\documentclass[printer]{aa} 

\makeatletter\let\expandableinput\@@input\makeatother

\usepackage{graphicx}
\usepackage{siunitx}
\usepackage{booktabs}
\usepackage{longtable}
\usepackage{lscape}
\usepackage{array}
\usepackage{xcolor}

\begin{document}

\title{Spatial and dynamical structure of the NGC\,2264 star-forming region}


\author{E. Flaccomio\inst{1} \and G. Micela\inst{1} \and G.
Peres\inst{1,2} \and S. Sciortino\inst{1}  \and E. Salvaggio\inst{3} \and L. Prisinzano\inst{1}  \and M. G. Guarcello\inst{1} \and L. Venuti\inst{4} \and R. Bonito\inst{1} \and I. Pillitteri\inst{1}}
\institute{
  INAF - Osservatorio Astronomico di Palermo, Piazza del Parlamento 1,  90134, Palermo, Italy
  \and 
  Dipartimento di Fisica e Chimica - Specola Universitaria - Università degli Studi di Palermo, Piazza del Parlamento 1, 90134 Palermo, Italy
  \and 
  Light, Nanomaterials, Nanotechnologies (L2n) Laboratory, CNRS EMR 7004, Université de Technologie de Troyes, 12 rue Marie Curie, 10004, Troyes Cedex, France
  \and
  SETI Institute, 339 Bernardo Ave, Suite 200, Mountain View, CA 94043, US}

\date{Received/Accepted}

\abstract
{The formation of stars within molecular clouds and the early stages of
stellar evolution (e.g., mass accretion and disk dispersal) are all active research
topics. The target of this study, NGC\,2264, is a benchmark star-forming region
in which these issues can be profitably studied.}
{We revisit the structure, dynamics, and star-forming history of NGC\,2264 
in order to advance our understanding of the
processes that lead from molecular clouds to protostars, stellar associations,
and the evolution of both.}
{We assembled a new extensive sample of NGC\,2264 members. To this end we used new
X-ray data obtained with the {\em XMM-Newton} telescope, GAIA eDR3 data, and an
extensive collection of public and published catalogs. Following a previous
suggestion that the star-forming region might extend significantly beyond
the better studied areas, our search covers a wide 2.5$\times$2.5 degrees
region in the sky.}
{Our catalog comprises more than 2200 candidate members, which is a $\sim$100\% increase over
previous determinations. We analyze their
spatial distribution and define new substructures. Using GAIA
parallaxes we estimate a new average distance to NGC\,2264 of 722$\pm$2\,pc and suggest that the
embedded Spokes subregion is $\sim$20\,pc farther away within the molecular
cloud. A complex dynamics is unveiled by the available proper motions and
radial velocities: we observe signs of global expansion and rotation. At the same time, we
observe the collapse and coalescence of two substructures in a region where active
star formation is taking place. The fraction of stars with disks and of those
undergoing circumstellar accretion varies significantly across the field, suggesting that star formation has been occurring for several million years. 
A particularly low accretion disk fraction around the O\,VII star S\,Mon might be attributed to external disk photoevaporation
or to an older age of the stars in the region.}
{NGC\,2264 is not dynamically relaxed and its present configuration is the result
of multiple dynamical processes. The cloud has been forming stars for several million years and we identify the process that is likely
responsible for the ongoing formation activity.}
   \maketitle
%

\section{Introduction}

Star formation (SF) begins with the gravitational collapse of giant molecular clouds
into molecular cores \citep{mac04}, which eventually evolve into protostars
through mass accretion onto a central object. Once the main accretion phase is
over, we are left with pre-main sequence (PMS) stars surrounded by circumstellar
disks \citep{bou07a}. These disks then dissipate within a few million years, as
the stars continue their contraction toward the main sequence (MS).
Understanding the mechanisms that drive the initial collapse and that regulate
the SF process requires the combination of theoretical efforts with
detailed observations of star-forming regions (SFRs) at different evolutionary
stages and in different environments \citep{lad03}. Indeed, the precise nature
of the involved physical and dynamical processes have yet to be clarified. For
example, it is still unclear whether SF is a slow, quasi-static
process, regulated by the magnetic field and ambipolar diffusion \citep{tas04},
or whether it occurs on shorter timescales, as expected if controlled by turbulence
\citep{bal07}. Furthermore, recent studies suggest that, rather than a single
SF episode, a molecular cloud may undergo several successive
formation bursts, forming hierarchical spatial and dynamical structures, before
becoming gravitationally relaxed. Also, the role of filaments is currently being
actively investigated as the preferred SF locus
\citep[e.g.,][]{bal87,and14,mon19b,sch20,nar22}. One approach to obtain a
clearer observational picture of SF is to try to reconstruct the
SF history of stars in active SFRs and to trace their stellar and
cloud dynamics. However, for unbiased results, the full spatial structure of the
region must be considered, including the outer regions often overlooked because
of observational constraints. We may thus assess the presence of subclusters and
trace them to temporally distinct and-or related SF episodes, and
follow the dynamic evolution of the structures, such as the merging of
filaments. Low-density haloes around SFR are also of great interest to obtain a
full picture of SF. They might, for example, trace the evaporation
of low-mass stars (leading to mass segregation), and thus be characterized by
low stellar masses and characteristic space velocities, or they might trace and
reveal older and now slowly expanding populations. Finding and characterizing
these extended and low-density populations can be challenging due to the large
contamination from field stars in the outskirts of SFRs. The problem is even
more severe if most of these stars are old enough to have dissipated their
circumstellar disks, making tracers of disks and accretion particularly biased
and incomplete. The present investigation aims at improving our understanding of
the spatial structure, SF history, and dynamical evolution of
NGC\,2264, which is one of the most prominent SFRs in the solar neighborhood. 

\begin{figure*}[!ht]
  \resizebox{\hsize}{!}{\includegraphics{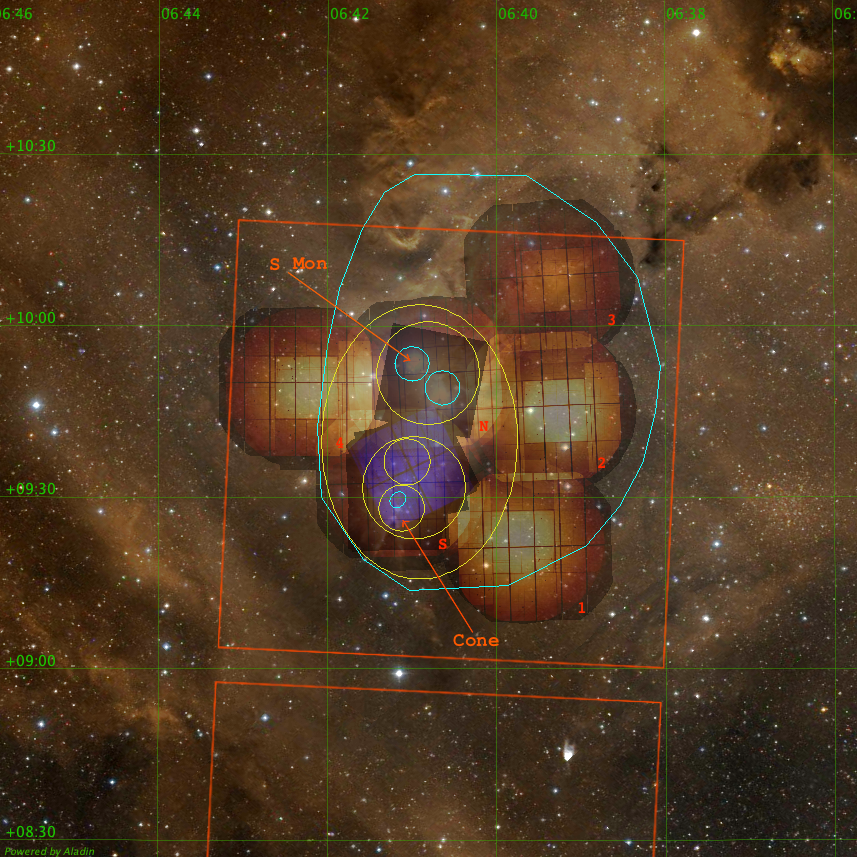}}
  \caption{Optical DSS2 image of the 2.5$\times$2.5\,deg square field centered on
  NGC\,2264 to which our study is limited. The locations of two prominent features, the O7 star S\,Mon and the Cone Nebula, are indicated by orange arrows.
  The areas shaded in blue and orange-brown are combined exposure maps of all
  available {\em Chandra} and {\em XMM-Newton} observations, respectively.
  Identifiers for the {\em XMM-Newton} field are also shown by red symbols in
  their respective lower-right corner. The two orange squares indicate the FoV of the  CoRoT observations discussed in Sect.\,\ref{sect:corotvar}.
  The yellow circles and ellipse indicate the
  subregions identified by \citet{sun09}. Contours of the four new spatial substructures identified in this study (Sect.\,\ref{sect:ResSpatialAndCont}), three inner ones and the outer polygon, are drawn in cyan.}
  \label{fig:FOV}
\end{figure*}

NGC\,2264 \citep{dah08} is a young stellar cluster populated by more than a
thousand stellar-mass members; it is located about 760\,pc from the Solar System
\citep{par00} and SF activity is still ongoing within its parental
cloud. The median age of the cluster is approximately 3\,Myr with an age
dispersion of at least 5 Myr \citep{dah08,ven19}. Thanks to its relative
proximity and low foreground extinction, NGC\,2264 is one of the most accessible
SFRs in our Galaxy. The MS and PMS populations of NGC\,2264 are
thus well-characterized, thanks to a large number of observations at all
wavelengths. Infrared and X-ray observations have identified a complex spatial
structure and two prominent sites of current SF activity in the
central areas of the cluster \citep[see, for example,][]{per06a}. \citet{sun08} have shown
that the SFR is preferentially elongated in the north-south direction and
extends at least up to 3-5 pc from the field center or 2-3 pc from the S\,Mon
O\,VII star and the Cone Nebula (see Fig.\,\ref{fig:FOV}). \citet{ven18} further
investigated the structure and SF history of NGC\,2264, using
spectroscopic and photometric data from the GAIA-ESO and CSI-NGC\,2264 surveys
\citep{ran13,gil12,cod14a}. They selected 655 likely members based on Li-,
H$\alpha$-, IR- and UV- excesses, and on X-ray emission, confirming its
hierarchical structure and an age spread of 4-5 Myr. Isochronal ages, gravity
indexes, disk and accretion fractions, indicated multiple SF events:
stars in the outer regions appear to be more evolved, having possibly been
formed earlier and migrated outward. Moreover, SF was found to have
started in the north and then propagated to the south, where it is still
ongoing. An age difference between disk-bearing and accreting stars was also
detected, indicating that disks outlast the end of the disk-accretion process.
Disks around the O-type star S\,Mon were found to be shorter-lived than in the
south possibly due to external photo-evaporation by the hot star.

GAIA DR2 data was employed by \citet{kuh19} to investigate the internal dynamics
of several young regions, including NGC\,2264. Unlike most of the other regions,
NGC\,2264 showed no evidence of expansion or of subcluster coalescence.
\citet{buc20}, focusing on the spatial clustering and dynamics of the central
regions of NGC\,2264, found evidence of a prolonged SF activity.
Also the clustering of stars was found to be  related to the evolutionary stage,
the least evolved being more clustered. This implies that stars form in compact
regions and then disperse. The compact subcluster of preferentially Class\,III
stars around S\,Mon, was taken as evidence of a significant stellar feedback
(photo-evaporation), confirming the finding of \citet{ven18}.

All of the studies described above were limited to the ``classical'' boundaries of
the star-forming region, that is the ``Halo'' defined by \citet{sun08} and indicated
in Fig.\,\ref{fig:FOV} by the outer yellow ellipse. \citet{mon19a} and
\citet{mon19b} focused on G02.3+2.5, a north-south filamentary region,
$\sim$0.2-1.0 degree north of the O7 star S\,Mon and studied its gas and dust
emission and dynamics. Their multiscale analysis, from the compact sources, to
cores and filaments, indicate that the recent and ongoing SF in
this region is likely due to the collision of two filamentary structures. The
collision itself might be due to the northward expansion of the H\,II region or
gravitational collapse toward the main cluster to the south. The wider-field
objective-prism H$_\alpha$ survey of \citet{rei04a} uncovered a small number of
dispersed accreting members, poorly correlated with the gas cloud and which
they therefore suggested might have already drifted away from their formation
locus. \citet{ven19} also surveyed a broad region. They were able to identify
the late-type population with limited field contamination, using exclusively
Pan-STARRS1 and UKIDSS photometry. They retrieved the spatial structures of
\citet{sun09}, but also identified a more widespread population of $\sim$100
stars up to 10-15\,pc away from the cluster center. As described below, clear
indications of a larger extent of the region have been also derived independently by
us, on the basis of the optical variability of $\sim$8000 stars in the NGC\,2264
field, obtained with the \textit{CoRoT} satellite \citep{bag06}. This
statistical study, briefly described in \S\,\ref{sect:corotvar}, was indeed our
initial motivation for obtaining new X-ray observations of the outskirts of
NGC\,2264 so to identify and characterize its dispersed population, eventually
aiming at a better understanding of how SF proceeded in the region.
Since YSOs have X-ray luminosities 2-4 orders of magnitude higher than field
stars \citep{wal88}, X-ray observations can easily distinguish high-probability
YSO candidates from the field stars. Furthermore, X-ray studies are
complementary to IR photometry: while this latter allows the identification of
YSOs with excesses due to the presence of disks and envelopes, it does not
identify PMS stars that have dissipated their disks.

In this work we also take advantage of a large body of literature data and, in
particular, of the GAIA early 3rd data release (eDR3), both for defining
membership and for characterizing young stars. As a byproduct of this effort we
also test our approach for selecting candidate members of a SFR on the basis of
optical variability. This approach might indeed prove useful in view of the
present and future availability of suitable and extremely large databases of
optical time-series for huge number of stars (such as Kepler, TESS, Rubin-LSST,
GAIA, etc..).

The article is structured as follows: in Sect. \ref{sect:corotvar} we give more
details about \textit{CoRoT} observations. In Sect. \ref{sect:obsdata} we
present literature data and the original X-ray observations we have used. For
the X-ray observations, we describe the data reduction, source detection, and
cross-identification procedures used to identify the optical and IR counterparts
of X-ray sources. Section \ref{sect:results} introduces our primary results: the
identification of samples of candidate cluster members, complete with
estimations of their field-star contamination, and the definition of new spatial
structures. In Sect. \ref{sect:discussion} we make use of our new member catalog
to address a number of issues: first we discuss the efficiency of member
selection via X-ray and variability criteria (\S\,\ref{sect:xvarsel}); we then
discuss the average physical properties of stars and their spatial variation.
More specifically we asses density, distances, kinematics, masses and ages, and
disk fractions. Finally, Sect.\,\ref{sect:conclusions} summarizes our work and
its main conclusions.

\section{Motivation - Optical variability with CoRoT}
\label{sect:corotvar}

CoRoT was a European satellite devoted to continuous optical observations of
star fields for asteroseismological and planetary transit studies. We observed
NGC\,2264 with the exo-planetary camera of CoRoT in 2008 with an
``additional program'' (PI: F. Favata), obtaining 23.5 days of accurate and
uninterrupted optical photometry for about 8000 preselected stars in a two
adjacent 1.7 sq.degree fields of view (FoVs), each imaged by a different CCD
(see Fig.\,\ref{fig:FOV}). The northern region encompassed the whole SFR, while
the southern one did not contain any members of NGC\,2264 known at the time of
the observation. The CoRoT targets were selected to include $\sim$270 known
NGC\,2264 members down to I$\sim$16, or M=0.3-0.4M$_\odot$ (see
\citealp{ale10}, \citealp{fla10}), but the vast majority were randomly selected
field stars in the same magnitude range. Their lightcurves are, as expected,
statistically much less variable than those of known NGC\,2264 members. However,
some of these ``field stars'' appeared to be as variable as members. We thus
embarked in a systematic search for variable stars that might be previously
unrecognized members of NGC\,2264. Variable field star contaminants were taken
into account adopting, as a control field, the FOV of the CCD pointed to the south
of NGC\,2264, which is certainly dominated by field stars\footnote{We also
obtained another CoRoT observation of NGC\,2264 in 2011 \citep{cod14a,fla18},
but it is not discussed here since only one of the exo-planetary CCDs
was available at that time.}.

For each CoRoT lightcurve we measured: $i)$ the standard deviation of the flux,
and $ii)$ the height of the maximum peak in the Lomb-Scargle Normalized
Periodogram (LNP, \citealp{sca82}). The standard deviation, divided by the
median flux, is a measure of the variability amplitude; the maximum of the LNP
is instead related to the degree of periodicity, irrespective of the period.
These two quantities are meant to preferentially single out Classical and
Weak-line T Tauri stars (CTTS and WTTS), respectively. Most CTTS, indeed, have
irregular or quasi-periodic lightcurves with large amplitudes, while WTTSs show
regular modulations and smaller amplitudes (\citealp{ale10}, \citealp{cod14a}).

\begin{figure}[!t!]
\centering
\includegraphics[width=9.0cm]{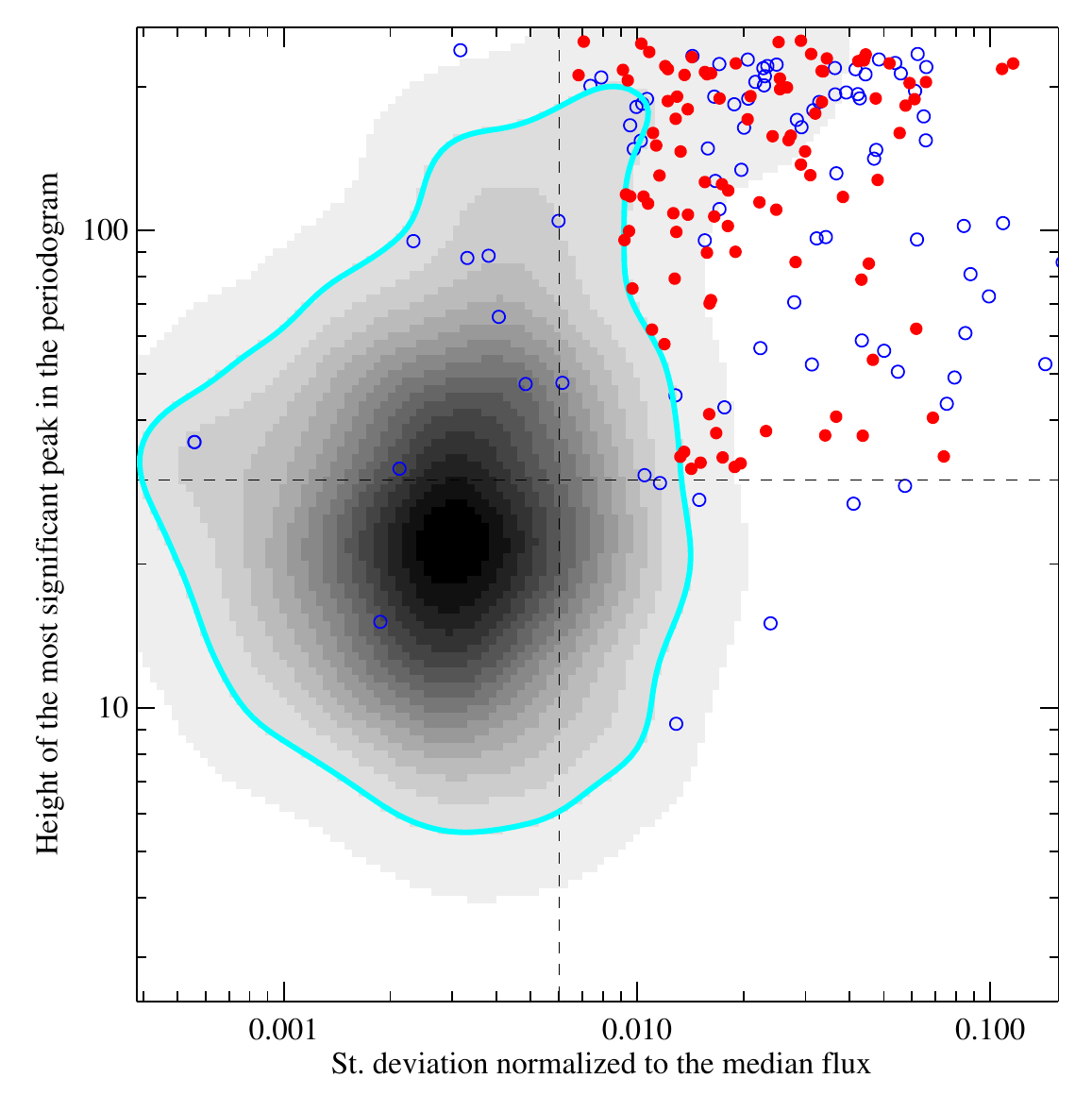}
\caption{Height of the most significant LNP peak vs. standard deviation of light
curves normalized to the median flux value. The grayscale map represents the
density of stars in the control field, with the cyan contour encircling 90\% of
these stars. Blue open points are known members of NGC\,2264, while red points
indicate the new candidate members.}
\label{fig:corot}
\vspace*{-0.3cm}
\end{figure}

We plot the two quantities derived from the CoRoT lightcurves\footnote{Because
of the large windows from which CoRoT magnitudes are estimated, contamination
from nearby stars is frequent. We therefore limited our analysis to isolated
stars with ``secure'' identifications, severely reducing our reference
samples.}, one versus the other, in Fig.\,\ref{fig:corot}. Empty blue circles
indicate a sample of likely cluster members, selected using literature data and
at least one of the following criteria: X-ray detection, mIR excess, strong
emission in the H$_\alpha$ line (from low-resolution spectroscopy or narrow-band
photometry). Members clearly cluster in the upper-right corner,  having
large-amplitude and periodic lightcurves. The locus of field stars, from
$\sim $3600 stars in the control field, is indicated by the grayscale map,
which is a smoothed density map (black stands for maximum density). The cyan
contour encircles 90\% of field stars (in the smoothed map, 92\% of actual
data-points). Seventy-four per cent of the 93 known members lie outside of this
contour and in the upper-right quadrant. We thus considered as candidate new
members all the stars in the northern CoRoT field that fall in this region of
Fig. \ref{fig:corot} and that were not previously known as members (filled red
circles). This selection results in a total of 170 candidate members (including
previously known ones). Notably, the new candidate members have a broader
spatial distribution with respect to the known members, and a noticeable higher
concentration toward the west of the cluster. On the basis of our control field,
 however, 3.2\% of the field stars are estimated to fall in the
member-selection region of Fig.\,\ref{fig:corot}. If we make the simplifying
assumption that a negligible fraction of new members is present among the 2089
stars in the north field that are not known members, we estimate $\sim$67 field
star contaminants (3.2\% of 2089), and a contamination fraction of 39\%
(67/170)\footnote{The approximation can be easily lifted: assuming that 74\% of
new members fall in the member-selection region of Fig.\,\ref{fig:corot}, we
estimate a very similar field contamination fraction of 38\%.}. This fraction
might actually be lower since the CCD covering NGC\,2264 also includes the
densest part of the molecular cloud and is thus expected to contain fewer
background stars than the control field, down to any given magnitude. In
Sect.\,\ref{sect:xvarsel} we reexamine the variability-selected sample to
assess the effectiveness of the chosen variability-based membership criterion.

\section{Observations and data analysis}
\label{sect:obsdata}

Stimulated by the CoRoT observation discussed in the previous section, we
obtained new X-ray data of the outskirt of the region, mainly to the west of the
known cluster area. X-ray detection will indeed serve as one of our primary
membership criteria in the outskirts of the region. We will also extensively
exploit, for the same purpose, the GAIA astrometric data, from the early data
release 3 (GAIA-eDR3). Additional optical and IR data from the literature will
also be used, both as help in establishing membership, and to subsequently
characterize the selected members. In the following two subsections we will
discuss the literature data and the new X-ray data, respectively. 

\subsection{Literature data}
\label{ref:litdata}

We collected a large body of available data for the region, in order to define
membership and characterize NGC\,2264 stars. The largest field that we consider
is a 2.5$\times$2.5\,deg square centered on RA=06:40:48, Dec.= 9:42:00, which likely
contains the whole NGC\,2264 population. An optical DSS2 image of the region is
shown in Fig.\,\ref{fig:FOV}. The yellow circles and ellipse toward the center
of the field delimit the subregions defined by \citet{sun09}. The outer
ellipse, in particular, is indicated by \citet{sun09} as the boundary of the
cluster ``Halo''. The four subregions drawn in cyan are newly introduced in
this paper (Sect. \ref{sect:ResSpatialAndCont}). We began with the master
catalog of cross-identified objects also used by \citet{cod14a}. This contains
several photometric and spectroscopic catalogs, optical and IR, mostly limited
to the inner part of the region. Specifically, we include: SDSS system
photometry from \citet{ven14}, optical Johnson-Cousins photometry
\citep{dah05,lam04,reb02,sun08}, UKIRT nIR photometry \citep{kin13}, Spitzer mIR
photometry \citep{sun08,cod14a}, rotational periods \citep{lam04,reb02,ven17},
and spectral types \citep{wal56,mak04,dah05}. To this catalog we added data from
the GAIA-ESO survey (DR5) and the following wide area surveys covering the full
2.5$\times$2.5\,deg field in Fig.\,\ref{fig:FOV}: GAIA (early 3rd data release,
\citealt{gai21}), Pan-STARRS1 (1st data release, \citealt{cha16,fle20}, PS1
hereafter\footnote{Following \citet{ven19}, the PS1 catalog was limited to
objects: i) with estimates for both $r$ and $i$ magnitudes, ii) brighter than
$r$=23.2 and $i$=23.1.}), IPHAS (2nd data release, \citealt{bar14}), 2MASS
(point source catalog, \citealt{skr06}), and allWISE \citep{wri10}. 

Each of these additional catalogs was added in sequence. Objects in our master
catalog were matched to those in each of the catalogs using identification radii
based on the provided positional uncertainties (limited to a minimum radius,
chosen for each catalog) and the spatial resolution of the survey. A thorough
visual examination was then performed at each step to resolve uncertain and-or
ambiguous identifications. The allWISE catalog received an ad-hoc treatment.
Because of the poor spatial resolution of the WISE mIR images, the appropriate
identification radii were rather large and visual inspection of the  $\sim1.3
\times 10^5$ catalog entries in our FOV turned out impractical. We therefore
decided to limit identifications to isolated cases, that is allWISE sources with
a single counterpart in our master catalog and for which the second closest
object was farther away than both 3 times the identification radius and five
arcseconds. A total of 7.6$\times10^4$ allWISE sources ($\sim$57\% of the full
catalog) where thus matched with our catalog. Following a preliminary membership
analysis (Sect.\,\ref{sect:membership}) we also identified the remaining allWISE
sources with all candidate members (member sample ``a'' in
Sect.\,\ref{sect:membership}). Following visual inspection, 2376 additional
allWISE sources were matched with 3151 stars in our master catalog. This latter
step was helpful both to assign mIR photometry to our candidate members and to
confirm the membership status of stars on the basis of their mIR excesses.

NGC\,2264 has been observed in X-rays multiple times with both {\em Chandra} and
{\em XMM-Newton}. Figure \ref{fig:FOV} shows the spatial coverage of these X-ray
observations. Four different sets of {\em Chandra} observations performed with
the ACIS-I detector, have been presented by \citet{ram04}, \citet{fla06},
\citet{fla10}, and \citet[][see also \citealt{fla18}]{cod14a}, covering the
central S\,Mon and Cone regions with total exposure time between $\sim$50\,ks
(toward S\,Mon) and $\sim$460\,ks (north of the Cone Nebula).

Most of the {\em Chandra} fields cover the southern cores, that is the Cone
Nebula, IRS1, and IRS2 (or Spoke-cluster) regions (PIs: Sciortino, Micela),
while a single shallower exposure (PI: Stauffer) covers the S\,Mon region. They
all fall inside the Halo region as defined by \citet{sun08,sun09}, the outermost
yellow ellipse in Fig.\,\ref{fig:FOV}. An analysis of all the above {\em
Chandra} observations was presented by \citet{tow19}, who found 3373 point
sources. However, we here prefer to use our own more conservative analysis of
the same data, to be fully described in Flaccomio et al. (in prep.). The
resulting catalog of 1043 X-ray sources was included in our master catalog.
Basic information on these sources is listed in Table\,\ref{tab:acis12_src}.

Finally, results for the two central {\em XMM-Newton} fields were published by
\citet{dah07}, but we did not include the respective source lists in our master
catalog since we reanalyze these observations consistently with the new ones
here.

\subsection{{\em XMM-Newton} X-ray data}

As mentioned above, two {\em XMM-Newton} exposures  (PI: Simon) were taken,
respectively, on 2001 March 20 and 2002 March 17, pointed toward the inner
regions of the cluster \citep{dah07}. They lasted $\sim$40\,ks and were
performed using a thick filter to minimize the contamination of the ultraviolet
light from the massive cluster stars. Hereafter, they are named field N (for
north) and field S (south). We obtained four additional observations between
2011 and 2014 each lasting about 50\,ks (P.I. E. Flaccomio). These
observations were pointed to the outer regions of the cluster and were performed
with a medium filter. Hereafter these fields are named field 1, 2, 3, and 4.
Figure\,\ref{fig:FOV} shows the footprints of the observed regions in the sky.
The first five columns in Table \ref{tab:obs} list, for each field, observation
IDs (in the {\em XMM-Newton} archive), pointing coordinates, and observation dates.

\subsubsection{Data reduction and source detection}

\begin{figure}[!h!]
  \resizebox{\hsize}{!}{\includegraphics{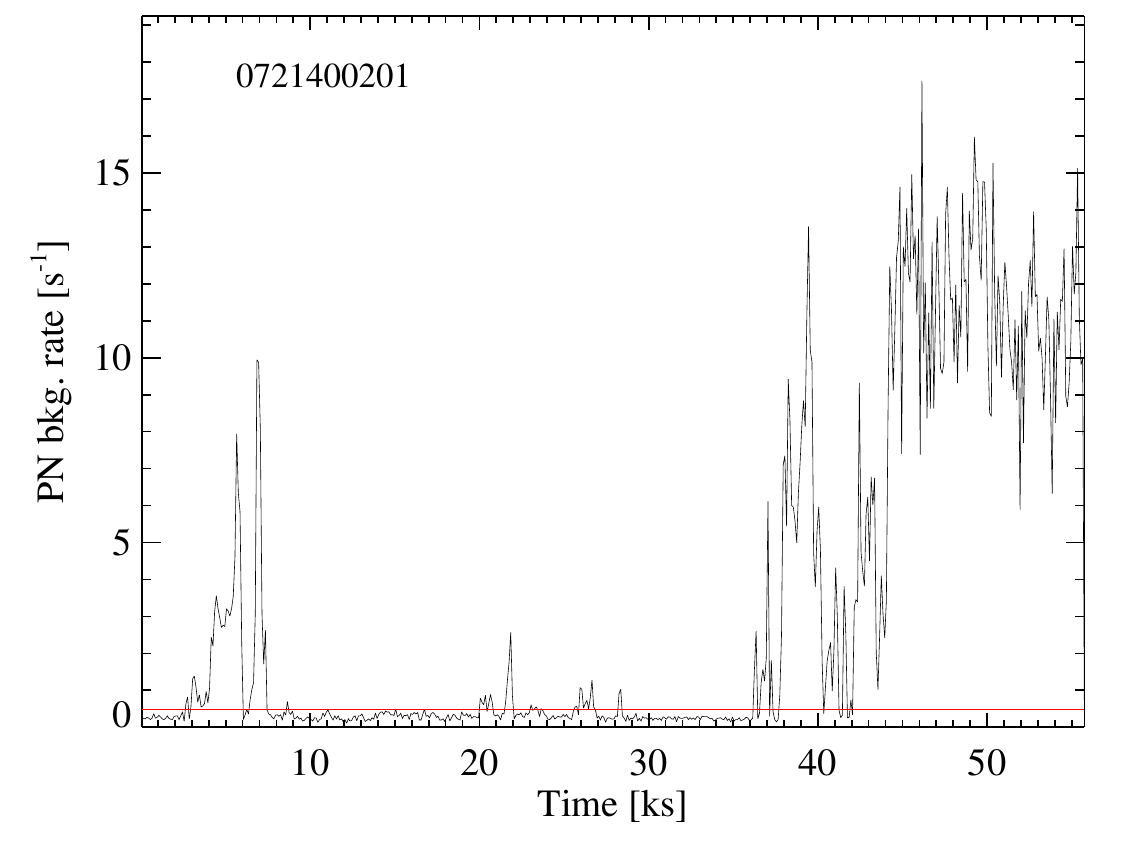}}
  \caption{Background light curve of EPIC PN-detected events in field 3 (Obs.ID 0721400201); the
  horizontal red line is the count-rate threshold that we chose to define and
  filter-out high-background time intervals, thus maximizing the S/N of faint
  sources.}
  \label{fig:bkglc}
  \end{figure}

The data of each \textit{EPIC} detector, MOS1, MOS2, and PN, were reduced with
the \textit{XMM–Newton} SCIENTIFIC ANALYSIS SYSTEM (SAS; \citealp{gab04})
version 14.0.0. We reprocessed the \textit{PN} and the \textit{MOS} data to
obtain tables of photons with astrometry, energies and arrival times. We then
filtered out artifacts due to bad pixels, bad columns and cosmic rays. In order
to  maximize the signal-to-noise ratio of weak sources, background levels during
each observation were monitored and data taken at times of high background were
removed. To this effect we followed the standard SAS data analysis
threads\footnote{\url{https://www.cosmos.esa.int/web/xmm-newton/sas- threads}}.
Figure\,\ref{fig:bkglc} shows, as an example, the lightcurve of background
photons detected by the \textit{EPIC-pn} for field 3, the one with the worst
background flares and, thus, the largest fraction of rejected data. The
lightcurve refers to events with energy between 10 and 12\,keV, pattern=0, and
standard PN flag filtering. In this case, time intervals with background count
rates greater than 0.5 cnt/s were discarded. This rejection threshold was chosen
individually for each observation and \textit{EPIC} camera
(\textit{PN},\textit{MOS1}, and \textit{MOS2}), by inspecting the relative
background lightcurves. Columns 6-8 in Table\,\ref{tab:obs} list the total and
background-filtered exposure times (on-times) for each observation and detector. 

\begin{table*}
  \small
  \caption{\textit{XMM-Newton} observations of NGC\,2264 - basic data, background filtering, and source detection\label{tab:obs}}
  \centering
  \begin{tabular}{l c c c c c c c r r}
  \hline\hline
  Fld & Obs.ID     &     R.A.   & Dec.        &   obs. date   &  $MOS1[t'/T]$\tablefootmark{a} &$MOS2[t'/T]$\tablefootmark{a} & $PN[t'/T]$\tablefootmark{a}  &   det. th.\tablefootmark{b} &    n. srcs\tablefootmark{c}  \\
      &            &  [J2000]   & [J2000]     &               &          [ks] &        [ks] &     [ks]    & [$\sigma$] &             \\
  \hline 
   N  & 0011420101 & 6:40:43.30 &  9:51:00.00 &  2001-03-20   &     40.7/38.1 &   40.7/38.2 &  36.3/33.5  &        4.9 &       251   \\
   S  & 0011420201 & 6:41:05.50 &  9:31:40.00 &  2002-03-17   &     41.0/40.9 &   41.0/40.9 &  38.7/38.5  &        4.9 &       299   \\
   1  & 0673820101 & 6:39:40.40 &  9:21:58.10 &  2011-09-21   &     53.5/53.4 &   53.5/53.5 &  52.0/51.9  &        5.0 &       129   \\
   2  & 0694210101 & 6:39:18.70 &  9:45:13.00 &  2013-03-22   &     58.5/57.4 &   58.6/58.3 &  57.0/51.4  &        5.0 &       179   \\
   3  & 0721400201 & 6:39:19.20 & 10:08:05.00 &  2014-03-27   &     56.8/36.8 &   56.8/35.8 &  50.6/30.6  &        4.9 &        90   \\
   4  & 0721400301 & 6:42:15.00 &  9:49:20.00 &  2014-03-12   &     51.6/50.6 &   51.6/50.8 &  50.0/47.8  &        5.0 &       133   \\
  \hline
  \end{tabular}
  \tablefoot{
  \tablefoottext{a}{$t'$ and $T$ indicate exposure times before and after filtering for high background periods, respectively.}
  \tablefoottext{b}{Significance threshold used for source detection.}
  \tablefoottext{c}{Number of detected sources.}
  }
  \end{table*}

Source detection was performed for events in the 0.3-7.9\,keV energy band,
appropriate for coronal sources, using the PWXDetect wavelet-based code
developed at the INAF - Osservatorio Astronomico di Palermo by \citet{dam97}.
The code combines the \textit{MOS} and \textit{PN} data so to maximize the
detection efficiency for faint sources. Significance thresholds for source
detection were determined so to yield, on average, one spurious detection per
field due to background fluctuations. These thresholds were determined
individually for each of our six datasets, by simulating 100 blank
(background-only) fields and running the detection code on these fields. The
blank fields were simulated adopting the actual exposure map and number of
background events, this latter determined subtracting the number of source
events, from a preliminary detection, from the total number of detected events.
The final significance thresholds, 4.9 or 5.0\,$\sigma$, depending on field, are
listed in column 9 of Table\,\ref{tab:obs} followed, in col. 10, by the number
of detected X-ray sources (1081 in total). Detections were inspected
individually so to spot obviously spurious sources produced by residual
instrumental artifacts that were not filtered out by the data processing
described above. Thirty-five sources were then removed from the detection lists:
twenty-nine in field 2, four in field N, and two in field S.

We finally created a merged {\em XMM-Newton} source catalog, by
cross-identifying the six individual lists from each observation. Because of the
(small) overlap between the fields, some sources are indeed detected in more
than one observation. We employed an iterative merging process: we started by
cross-identifying sources detected in field N and field S. The resulting source
list was then cross-identified with that of field 1, and so on until all lists
were merged. The cross-identification radii of individual detections were
arbitrarily chosen as $\sqrt{2}$ times the formal positional uncertainty
provided by the PWXDetect code, which is based on the observed count statistic
and photon spatial distribution. After merging each catalog, new positions and
uncertainties were computed as the uncertainty-weighed averages of those of the
matched detections. After each step, all identifications were visually checked
and compared with the spatial distribution of photons. In eight cases we
manually cross-identified source pairs that had not fulfilled our matching
criterion. This can be quite reasonable, since our identification radii, based
on positional uncertainties, can be significantly smaller than the spatial
extent of X-ray sources (for example for extended sources or, more often in our
case, groups of unresolved point sources). Our final catalog of unique {\em
XMM-Newton} X-ray sources, 944 entries, is presented in
Table\,\ref{tab:xmm19_src}. We list: sky positions, uncertainty-based
identification radii, detection significances, background-subtracted source
counts, and photon fluxes, as estimated by the PWXDetect code\footnote{For
sources detected in more than one observations we report the combined values for
identification radii and significances, the total number of source counts, and
the uncertainty weighted average fluxes.}.

\begin{table*}
\begin{center}
\caption{Detected {\em XMM-Newton} sources (full table available on-line)\label{tab:xmm19_src}}
\begin{tabular}{r r r r r r S[table-format=3.0(2)]}
\hline\hline
\expandableinput xmm19_tab.tex 
\hline
\end{tabular}
\tablefoot{
\tablefoottext{a}{Radius used for identification with master catalog.}
\tablefoottext{b}{Significance of detection.}
\tablefoottext{c}{Background-subtracted source counts.}
\tablefoottext{d}{Detected flux in the 0.3-7.9\,keV band.}
}
\end{center}
\end{table*}

\subsubsection{Cross-identification with the master catalog}

We matched the X-ray sources with our master catalog, described in
\S\,\ref{ref:litdata}, using a positional criterion. Identification radii were
chosen as the quadrature sum of the identification radii for the {\em
XMM-Newton} sources and those for the master catalog. For {\em XMM-Newton}
sources the radii are listed in Table\,\ref{tab:xmm19_src} and range between
1.0\arcsec and 10.7\arcsec (median: 3.6\arcsec). For the master catalog, radii
were taken, for each object, as the smallest positional uncertainty among those
adopted for the original catalogs in which the object is present. They range
between 0.3\arcsec and 8.1\arcsec (median: 0.5\arcsec).

A significant fractions of X-ray sources end up with multiple counterparts in
our master catalog. This is due, on one hand, to the sizable identification
radii for the X-ray sources, and on the other, to the large density of objects
in our master catalog, the large dynamic range covered, and, quite importantly
in parts of the FoV, the presence of many obviously spurious entries in some of
the constituent optical-IR catalogs. The IPHAS and UKIRT catalogs we have
adopted, in particular, clearly include a large number of spurious sources, many
of which can be ascribed to the nebulosity in the region or to the wings of
bright stars. For many {\em XMM-Newton} sources, therefore, some of the potential
optical-IR counterparts can be discarded with confidence. 

We thus resorted to inspecting each identification by eye and evaluating the
multiple optical-IR counterparts of any given X-ray source based on: differences
in magnitudes, in one or more bands, and in positional offsets from the X-ray
source, their position relation to the actual spatial distribution of X-ray
photons, deep optical images (such as those from Pan-STARRS), which help
identify likely spurious sources. In particular, we most often discarded
counterparts that were $>$5 magnitudes fainter than another counterpart. We
also often discarded faint counterparts only listed in the UKIRT or IPHAS
catalogs, when other brighter and more credible counterparts, often detected in
multiple optical-IR catalogs, were also present. At the end we discarded 1866
counterparts to 570 {\em XMM-Newton} sources, out of the initial 3050
counterparts to the 944 {\em XMM-Newton} sources.

In several cases, we also added counterparts that formally did not match the
positional identification criterion. In some cases, a prominent optical-IR
source fell close to the edge of the identification circle, whose size is based
on the positional uncertainty with a given confidence level. In these cases both
statistical and unaccounted-for systematic uncertainties (for example due to the
complicated PSF of {\em XMM-Newton}) can explain the observed offsets. In a
number of cases, two (or more), often similarly bright, optical-IR sources may
lie at opposite sides of the identification circle. Looking at the X-ray photons
spatial distributions, some of these X-ray sources are more extended than
expected for a point source. In some cases the X-ray source can even be seen to
be marginally resolved into two separate components. In these cases, in which
the uncertainty-based identification radii are not very informative, we added
all the reasonable counterparts outside of the identification circles. A total
of 129 counterparts were added for these reasons to 103 {\em XMM-Newton}
sources.

The result of the above changes to the initial identifications is that the
number of X-ray sources with unique identifications is increased from 200 to 680
and the number of X-ray sources with no optical-IR identification is reduced
from 108 to 75.

Finally, we inspected once more the 264 X-ray sources that, after the above
steps, were left with more than one possible counterpart. In doing so we
searched for additional optical-IR counterparts that were somewhat unlikely to
be responsible for the observed X-ray emission, for example because of their
larger offset and significantly smaller optical-IR luminosity with respect to
other counterparts. Contrary to the discarded counterparts described above,
these identifications were kept but were flagged as unlikely. To this effect we
examined again positional offsets from the X-ray position and magnitude
differences, and considered as unlikely counterparts those that had both
significantly larger offsets and were $>$2 magnitudes fainter than the brighter
and closer counterpart(s). We also examined the diagrams that provide membership
information and which are fully discussed in \S\,\ref{sect:membership}. In cases
in which at least one of the counterparts appeared to be an NGC\,2264 member, we
considered as unlikely counterparts those that had significant indication of
being nonmembers (field stars or background AGNs, statistically expected to be
fainter than members in X-rays). In a couple of cases we considered as unlikely a
counterpart which appear to be a spurious source in the original literature
catalog, based on deep archival optical images and other plausibility criteria
(for example extremely close pairs of sources that could not have been resolved in the
literature data). For {\em XMM-Newton} sources also detected at higher spatial
resolution by {\em Chandra}, we also considered as unlikely counterparts
optical-IR objects that were $>$2 mag fainter than the unique counterpart of the
{\em Chandra} source and that were clearly not detected in the {\em
Chandra} data. Disregarding all the above unlikely counterparts, the number of
{\em XMM-Newton} sources with a single likely counterpart increases from 680 to
735.

\section{Results}
\label{sect:results}

The master catalog we have assembled lists $\sim$6.5$\times 10^5$ objects
throughout our 2.5$\times$2.5\,deg FoV. We now proceed to use the information
gathered from the X-ray and optical-IR catalogs to identify members of
NGC\,2264, estimate field-object contamination, and discuss the spatial extent
and structure of the cluster.

\subsection{Candidate member selection}
\label{sect:membership}

\begin{figure*}
\centering
\resizebox{0.98\textwidth}{!}{\includegraphics{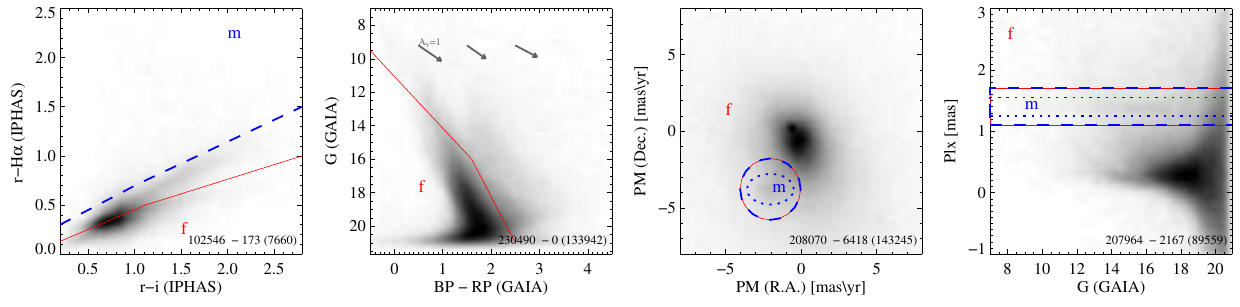}}
\resizebox{0.98\textwidth}{!}{\includegraphics{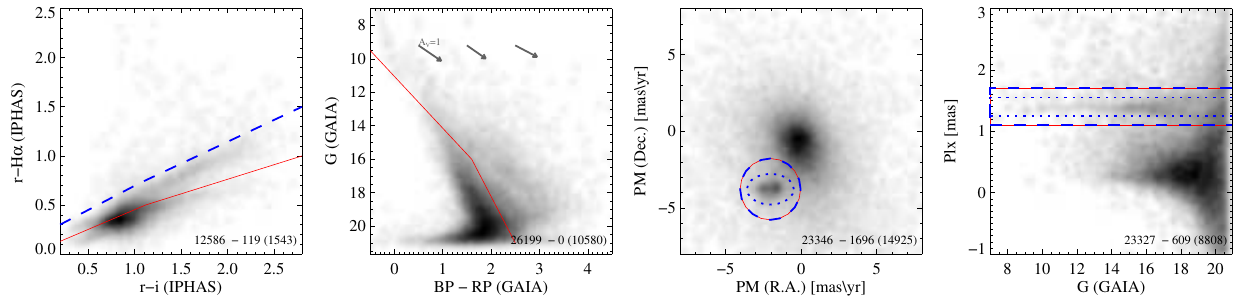}}
\resizebox{0.98\textwidth}{!}{\includegraphics{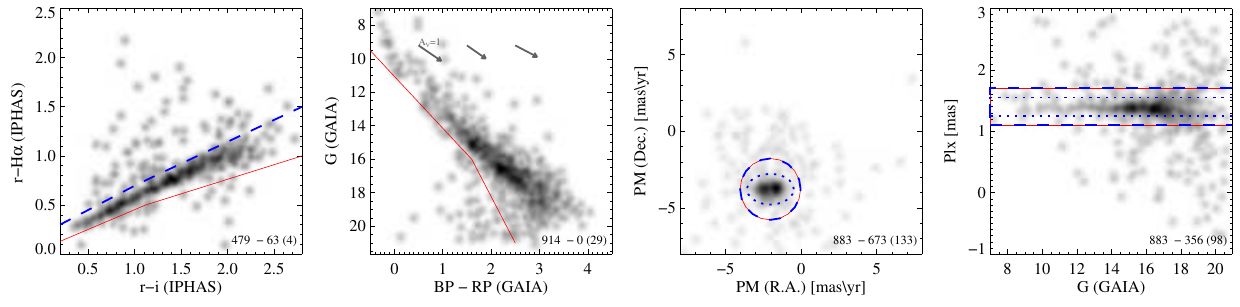}}
\resizebox{0.98\textwidth}{!}{\includegraphics{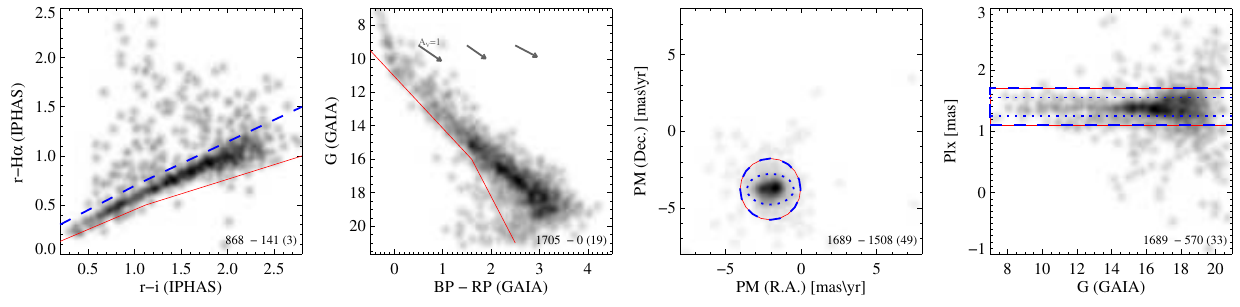}}
\caption{Object density in four different diagrams (one per column) and four
different samples (one per row). The diagrams are, from left to right:  the
r-H$_\alpha$ vs. r-i color-color diagram using IPHAS data, G vs. BP-RP color
magnitude diagrams (with three reddening vectors, at different colors),
PM$_{Dec}$ vs. PM$_{RA}$, and parallax vs. G magnitudes (the last three diagrams
using GAIA data). The four samples, from top to bottom, are: all objects in our
catalog, objects within the {\em XMM-Newton} field of view, objects
unambiguously detected in X-rays, candidate members in our sample ``c''
(Sect.\,\ref{sect:membership}). Density map are smoothed using a Gaussian Kernel.
The grayscales range from black
at maximum density to white at zero density, scaling with the square root of the
density, rather than linearly, to reveal more details at low-densities.
Isolated gray dots in low-density areas indicate individual stars.
Dashed and solid lines delineate member and field loci, respectively, and, in the first row, ``m'' and ``f'' labels help clarify the distinction between the two. Dotted lines in the last two
diagrams refer to more conservative member loci. The first number in the
bottom-right corners indicate the total number of plotted objects. The following
ones refer to the number of stars in the member- and field- loci, with 3- and
2-$\sigma$ significance, respectively.}
\label{fig:GAIAdia}
\end{figure*}

The first row of panels in Fig.\,\ref{fig:GAIAdia} shows, for all objects in our
catalog, the IPHAS r-H$_\alpha$ versus r-i color-color diagram, and three
GAIA-based diagrams utilizing: GAIA magnitude and colors (G versus
BP-RP)\footnote{Reddening vectors are shown in this and other color-magnitudes
and color-color diagrams presented in the following
(Fig.\,\ref{fig:diagrams_all}). For the broad GAIA bands, the reddening law was
derived, as a function of stellar temperature and intrinsic BP-RP color,
following the same procedures used in \citet{fla18} for the CoRoT passband. In
the present case we used the passbands reported in \citet{eva18} and the
\citet{wei03a} extinction law. As for photospheric models we used the ATLAS9
ones \citep{kur93a} for stars bluer than BP-RP=1.0 (T$_{eff}>$5400\,K) and the
BT\-Settl ones \citep{bar15} for redder stars. Three reddening vectors are
plotted in Fig.\,\ref{fig:GAIAdia}, referring to intrinsic BP-RP= 0.5, 1.5, and
2.5.}, the two components of the proper motion vector (PM$_{R.A.}$ versus
PM$_{Dec}$), and parallaxes (PLX versus G). Given the large number of plotted
points (first number in the bottom-right of each panel) we plot smoothed
grayscale maps of the point density. The maps are binned into 128x128 pixels and
convolved with a Gaussian kernel (FWHM=3\,px).  Dashed
blue lines in the plots indicate member loci, defined below, from which
candidate members are selected, with varying degree of contamination from
field objects (dotted blue lines, when shown, indicate more conservative,
less-contaminated, definitions). Solid red lines indicate the boundaries of
field loci, from which likely field objects are identified (in two of the
panels plotted here the boundaries of the field loci coincide with those of
member loci but this is generally not the case).
The second row of panels in Fig.\,\ref{fig:GAIAdia} shows
the same four plots for the subsample of objects falling within the XMM-Newton
field of view. Field object contamination within all defined member loci is
clearly severe for both the full sample and for this latter subsample. The
cluster can, however, be clearly identified in the proper motions and parallaxes
plots referring to the smaller field. The third row refers to X-ray sources,
detected by {\em Chandra} and-or {\em XMM-Newton}, and with an unique (or
preferred) optical-IR counterpart. Comparison with the preceding row clearly
demonstrates the power of X-ray observations in selecting candidate young
members of a star-forming region. Indeed, using these plots for X-ray sources,
we are able to define member and field-object loci for the full sample. Finally,
the fourth row of panels in Fig.\,\ref{fig:GAIAdia} refers to our working
candidate member sample (sample ``c''), which is introduced below.

In addition to the plots described above, using IPHAS and GAIA measurements,
several other diagrams providing membership information can be constructed with
our data. We used 19 additional plots, presented in
Appendix\,\ref{sect:modediagrams}. In 16 of these we defined field-star loci,
while in 9 we defined member-loci. As mentioned, for each of the panels in
Fig.\,\ref{fig:GAIAdia} and \ref{fig:diagrams_all}, we selected as candidate
members all the objects that fall in the member locus (delimited by dashed blue
lines), with a 3$\sigma$ confidence, where the confidence was estimated from the
uncertainties in the plotted quantities and Monte Carlo methods\footnote{A
two-dimensional Gaussian cloud of 10$^6$ points was created for each object and
the fraction of points falling within the locus was then converted to
significance.}. The same procedure was followed to identify likely field
objects, with the difference that the confidence level for this selection was
lowered to 2$\sigma$ (eventually resulting in a more aggressive vetting of
members).

We end-up with twelve membership criteria from these plots, plus X-ray
detection, and twenty-one exclusion (field object) criteria. Because of the
widely variable spatial coverage and depth of the diverse datasets we have
collected, only a fraction of the $\sim$6.5$\times 10^5$ objects in our full
catalog have the data to test each of these criteria. The numbers of objects
placed in each plot are given in the lower-right corner (first figure), and
range between 550 for the spectroscopic H$\alpha$ data and 3.4$\times 10^5$ for
the J versus J-H plot. A significant fraction of objects can be placed in one or
more of the plots providing membership criteria (68\%) or field-object criteria
(86\%). For 23\% (67\%) of the objects three or more membership (field-object)
criteria can be defined.

In both Fig.\,\ref{fig:GAIAdia} and \ref{fig:diagrams_all}, the numbers of
candidate members and field-objects selected by each criterion are shown by the
second and third figures in the lower-right corner of the relative plot. The
number of candidate members range between 173, for the IPHAS H$\alpha$ data, and
6418 for the PM data. Most, if not all, of the selected member subsamples
are contaminated by field-objects to varying degrees. For example, it is clear
from the PM plots that, over the whole field, candidate members selected from PM
alone are actually dominated by field objects belonging to the broad
distributions of points roughly centered at PM$_{R.A.}$=0, PM$_{Dec.}$=0.
Indeed, 79\% of these candidate members also fall in the field-object locus
(with 2$\sigma$ confidence) of at least one other plot. The same considerations
apply to the 2167 candidate members selected in the parallax versus G plane, 78\%
of which also fall in at least one field-object locus\footnote{Narrowing the
definition of the cluster loci in the PM or parallax-G planes, as shown in
Fig.\,\ref{fig:GAIAdia}, significantly reduces the number of selected candidate
members (by factors of 2.7 and 4.4), but the fraction of contaminants, as
deduced from the field-object loci, remains high, 58\% and 74\%, for PM and
parallaxes, respectively.}. Lower but significant ``contamination fractions''
(as approximated above), are also estimated for other membership criteria, with
the lowest estimates, 25\%, 28\%, 31\%, and 33\%, found for criteria based on
H$_\alpha$ EW, H$_\alpha$ FWZI, X-ray detection, and the {\em Spitzer}
[3.6]-[5.8] versus [3.6]-[4.5] color-color diagram, respectively.

Conversely, actual NGC\,2264 members are likely to fall into (and contaminate)
several field-object loci, for example because of the very large extinctions of
stars highly embedded in the molecular cloud, or large systematic errors in the
photometry in one or more bands (for example because of contamination from neighboring
objects or nebulosity), or even because of more physically interesting causes.
One example are the so-called below-main-sequence stars, which can easily be
mistaken for field stars. These are often found in young clusters and their
peculiar magnitudes or colors may be due to the edge-on viewing geometry of the
star-disk systems, and the associated attenuation and scattering of their
photospheric emission by the disk material \citep[see, for example,][]{gua10,bon13}.

Given the significant contamination of individual membership and field-object
criteria, we try to combine them to define samples of likely members that may
serve different purposes. Our sample ``a'' tries to include most members,
regardless of contamination; sample ``b'' tries to include only bona-fide
members, minimizing contamination; sample ``c'', finally, provides a reasonable
compromise between completeness and contamination.

Sample ``a'' includes the 10267 stars for which we have any indication of
membership, that is those that fall in at least one of the member loci or are X-ray
detected (with unique optical-IR identification). All of these stars are listed
in Table\,\ref{tab:candmemb}, available in its entirety on-line. For each of
these objects we present sky coordinates along with GAIA, PS1, 2MASS, CSI, {\em
Chandra}, and {\em XMM-Newton} identifiers. Two sequences of flags then show the
criteria indicating member and field-object status. In the following, we do not
further discuss sample ``a'', focusing instead on subsamples with smaller
contamination from field objects. 

After extensive experimentation, and guided by the estimates of field star
contamination discussed in the next section (\S\, \ref{sect:fied_cont}), we
define samples ``b'' and ``c'' as comprising stars that: (i) fall in more member
loci than field-object loci, or, (ii) fall in at least three member loci, or
(iii) are indicated as members by at least two of the ``stronger'' criteria,
that is loci indicative of accretion or presence of disks, and X-ray detection.
This latter provision has the advantage of being able to retrieve several of the
below-main-sequence members discussed above. Sample ``b'', which yielded the
lowest field star contamination fraction among the investigated samples, was
defined as the 1971 stars fulfilling the above criteria when considering the
most stringent member loci for PMs, parallaxes, and RVs, (the two rightmost
panels in Fig.\,\ref{fig:GAIAdia}, and fifth panel in
Fig.\,\ref{fig:diagrams_all}, respectively). Sample ``c'', finally, comprises
the 2257 stars that fulfill the same criteria when considering the larger, less
conservative, member loci. The last column in Table\,\ref{tab:candmemb} lists
the samples each star in sample `a' belongs to. Stars in sample `c' are plotted
in the fourth row of Fig.\,\ref{fig:GAIAdia} and in
Fig.\,\ref{fig:diagrams_sampc}.

\begin{table*}
\begin{center}
\caption{All candidate members (sample `a'): coordinates, identification with main catalogs, and membership flags\label{tab:candmemb}}
\resizebox{0.99\textwidth}{!}{
\begin{tabular}{lrrrrrrrrrrl}
\hline\hline
\expandableinput members_tab.tex
\hline
\end{tabular}
}
\scriptsize
\tablefoot{
\tablefoottext{a}{Membership flags - 2: in conservative locus with 3$\sigma$ confidence, 1: in standard locus with 3$\sigma$ confidence, 0: not in locus with 3$\sigma$ confidence, -: not placed in relative diagram. Flags refer to loci (or criteria) defined in the following planes, from left to right: r-Ha vs. r-i (IPHAS),
R-H$_{\alpha}$ vs. V-I (IPHAS),
R-H$_{\alpha}$ vs. V-I \citep{sun08},
R-H$_{\alpha}$ vs. R-I \citep{lam04}, 
H$_{\alpha}$(EW) vs. V-I,
H$_{\alpha}$(FWZI) vs. V-I \citep{bon20},
J-H vs. H-K,
[3.6]-[5.8] vs. [4.5]-[8.0]  (Spitzer),
W1-W2 vs. W2-W3 (WISE),
<RV> vs. G,
PM$_{Dec}$ vs. PM$_{RA}$,
PLX  vs. G,
i-z (PS1) vs. [3.6]-[4.5] (Spitzer),
X-ray detection (1/0/-: detected/not detected/not in FoV)\\}
\tablefoottext{b}{Field flags - 1: in standard locus with 2$\sigma$ confidence, 0: not in locus with 2$\sigma$ confidence, -: not placed in relative diagram. Flags refer to loci defined in the following planes, from left to right:
r-Ha vs. r-i (IPHAS),
R-Ha vs. V-I \citep{sun08},
R-Ha vs. R-I \citep{lam04}, 
H$_{\alpha}$(FWZI) vs. V-I \citep{bon20},
G vs. BP-RP,
I vs. V-I,
I vs. R-I,
J vs. J-H,
K vs. H-K,
J-H vs. H-K,
[3.6] vs. [3.6]-[4.5],
[8.0] vs. [4.5]-[8.0],
[24] vs. [4.5]-[8.0],
[3.6] vs. [3.6]-[24],
[3.6]-[5.8] vs. [4.5]-[8.0],
W1-W2 vs. W2-W3 (WISE),
<RV> vs. G,
PM$_{Dec}$ vs. PM$_{RA}$,
PLX  vs. G,
G - DM vs. BP-RP
r vs. r-i (PS1)\\}
\tablefoottext{c}{The sample(s) the stars belong to: `b', 'c', and `c-wide' (indicated by the letter `w'.)}
}
\end{center}
\end{table*}

Since most membership criteria can only be applied to a (small) fraction of the
considered field, in order to avoid spatial biases and, crucially, to discuss
the overall spatial distribution of stars in the region, we also introduce a
further sample, defined similarly to sample ``c'', but only using
information and catalogs available for the full field: 2MASS, GAIA, PS1, IPHAS, and
WISE. This latter sample includes 1828 star and we refer to it as
``c-wide'' (indicated as `w' in the last column of Table\,\ref{tab:candmemb}).
In spite of the better homogeneity across the plane of the sky, it is also
biased in several ways, for example because it includes a smaller fraction of the most embedded
(and youngest) members which are best selected through deep X-ray and {\em
Spitzer} mIR data.

\begin{figure*}
  \centering
  \resizebox{\textwidth}{!}{
  \includegraphics{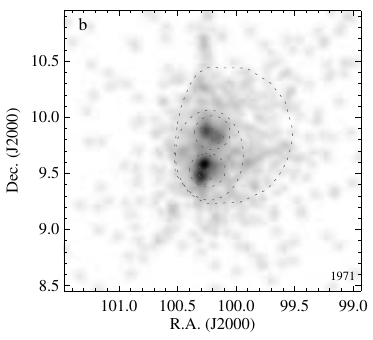}
  \includegraphics{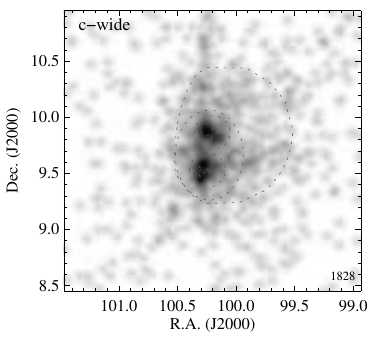}
  \includegraphics{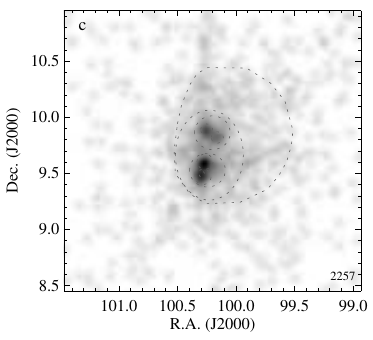}
  } \caption{Spatial distributions for candidate member samples ``b'', ``c-wide'', and
  ``c'' (from left to right). The density maps are smoothed with a Gaussian kernel
  as in Fig.\,\ref{fig:GAIAdia}, and dotted lines show the boundaries of the
  regions discussed in the text (Sect. \ref{sect:ResSpatialAndCont}) and also shown in Fig.\,\ref{fig:FOV}. The figures at the bottom right of each panel
  indicate the number of objects in the sample.} 
  \label{fig:radec_sm}
  \end{figure*}

Figure\,\ref{fig:radec_sm} shows sky density maps for samples ``b'', ``c-wide'',
and ``c''.

\subsection{Field-object contamination}
\label{sect:fied_cont}

\begin{figure*}
\centering
\resizebox{\textwidth}{!}{
\includegraphics{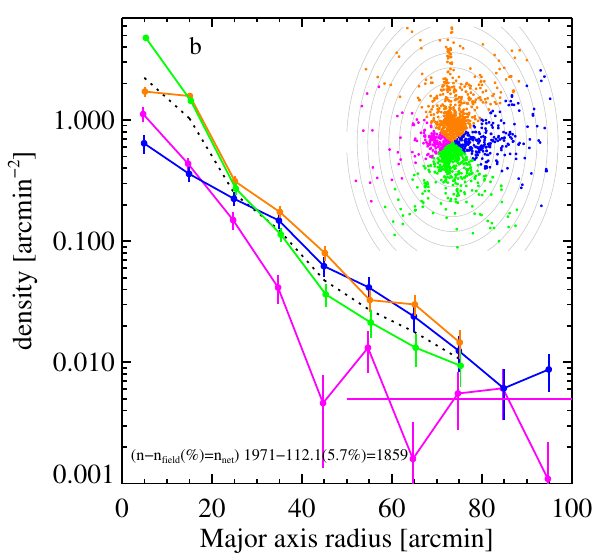}
\includegraphics{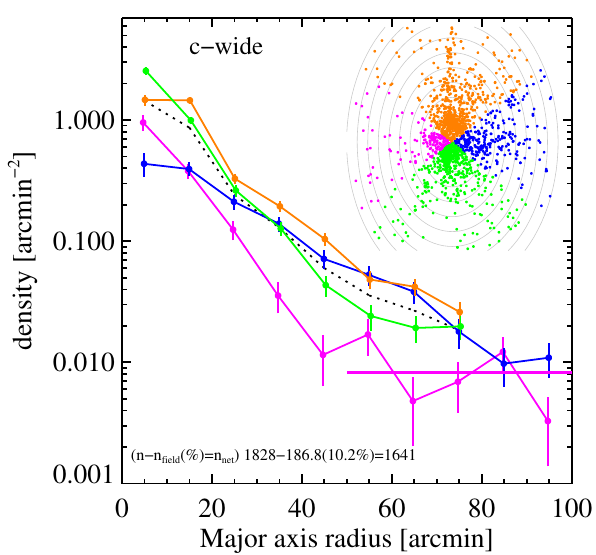}
\includegraphics{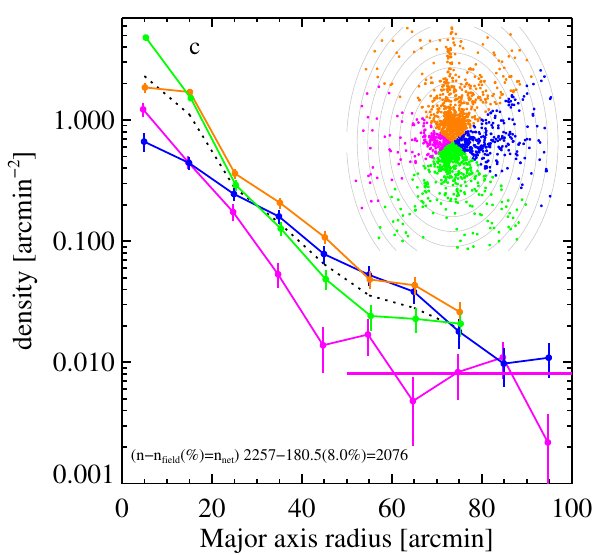}
} \caption{``Radial'' distributions for samples ``b'', ``c-wide'', and ``c''.
Different colors refer to different ``quadrants''. Spatial distributions are
shown at the top-right of each panel, with stars color-coded according to their
quadrant. The elliptical annuli where the densities for the main plot were
computed are also shown.  Points with error bars, based on Poisson
uncertainties, are slightly shifted along the x-axis to avoid overlap among the
four independent samples. The dotted black line refer to the combination of the
four quadrants. The thick horizontal magenta segments  toward the bottom right
indicate the average densities of stars in  the left quadrant for
r$>$50\,arcmin.}
\label{fig:field_cont}
\end{figure*}

Because of the complex procedure with which we built our candidate member
samples, estimating the degree of field-object contamination is not
straightforward. Assuming that the SFR is fully contained within our
2.5$\times$2.5\,deg field, we decided to estimate the sky density of field objects from
the density of member candidates far from the cluster center. More specifically,
we start by constructing density profiles of the candidate-member samples.
Considering the overall north-south elongation of the association and its
asymmetry with respect to the north-south direction, as apparent from
Fig.\,\ref{fig:radec_sm}, we construct our density profiles considering
elliptical annuli (rather than circular) and four separate quadrants (north,
south, east, and west). The elliptical annuli, in 10' steps, are centered and
have the same ellipticity (a/b=0.7) as the Halo region defined by \citet{sun08}.
Figure\,\ref{fig:field_cont} shows, in different panels, these profiles for
samples ``b'', ``c-wide'', and ``c''. The insets in the upper-right corners show
the spatial distribution of stars in each sample, color-coded according to the
quadrants, and superimposed on the elliptical annuli used to compute densities.
The main plots show the stellar density as a function of major axis radius (R),
separately for the four quadrants.

The density falls-off more rapidly in the eastern direction (magenta curve), as
also clearly observed in Fig.\,\ref{fig:radec_sm}, and appears to level off for
R$>$50'. We thus estimate the density of contaminants as the mean density of
objects in the eastern quadrant for R$>$50', indicated in
Fig.\,\ref{fig:field_cont} by a horizontal magenta segment. We report, at the bottom
of each panel, the number or stars in the sample, the estimated contaminants,
both in number (density $\times$ area of the full field) and as a percentage of
the total, and the estimated number of real members within the sample. We must
keep in mind, however, that these estimations are subject to several biases.
First of all, the actual density of field objects, at a given sensitivity, is
significantly greater in the outskirts rather than at the field center, due to
the obscuring cloud. Indeed, using the same procedure as above (plots not
shown), we find that: (a) the density of all GAIA-PS1-2MASS-WISE objects is
$\sim$2-3 times larger in the outer region than in the center\footnote{We do not
consider IPHAS-only sources because of the very large number of spurious
detections toward the field center} (b) the density of field objects, here taken
as all those that meet any of the criteria available for the full field, is
$\sim$5 times larger in the outer region wrt. the center of the field. This is
particularly relevant for sample c-wide, which, being based on full-field data,
is our simplest case. Because of the above, the contamination estimated from the
outer regions for sample c-wide is most likely an upper limit.

Samples ``b'' and ``c'', however, are selected using also datasets which are
limited to the field center, and with varying coverage. Indeed, the total
density of objects in our catalog is $\sim$7 times higher in the center than in
the outskirts\footnote{A significant fraction of objects in the center of the
field are, however, due to spurious UKIRT and IPHAS detections}. Consequently,
the number of both membership and field-object criteria used is greater in the
center of the field than in the outer regions, which is where we estimate the
density of contaminants. It is not easy to predict what the outcome in terms of
contaminating fraction is: while each additional membership criteria may
introduce some spurious member candidate, additional field-object criteria help
reduce contamination. Overall, however, the overwhelming majority of the
contaminants are expected to originate from the proper motion and parallax
criteria, based on GAIA data, which are available for the full field. Fewer
contaminants are actually expected from objects that are prominently missing
from the GAIA catalog, such as embedded objects (toward the field center)
selected as members through X-ray detection and-or mIR excesses. We thus argue
that our estimations of contaminants, ranging from 5.7\% for sample ``b'' to
10.2\% for sample ``c-wide'', are reasonable approximations and, possibly, upper
limits.

\subsection{Spatial extension and structure}
\label{sect:ResSpatialAndCont}

We are now in the position to discuss the extension of NGC\,2264 in the plane of
the sky, starting from the density maps for samples ``b'', ``c-wide'', and ``c''
shown in Fig.\,\ref{fig:radec_sm}. In spite of the differences in completeness
and-or contamination levels of the three samples, in all the maps we tentatively
identify a new extended structure, marked by the outer polygon (dotted).
Crucially, the region is visible even for the ``c-wide'' sample which, if
anything, is likely to have a bias against including embedded stars toward the
field center. We name this structure the ``Extended Halo'' to distinguish it
from the Halo, which is the smaller elliptical region within the polygon. We refer
to the region outside the extended halo, as the ``Field''.
 
\begin{figure}[!t]
\centering
\resizebox{0.4\textwidth}{!}{\includegraphics{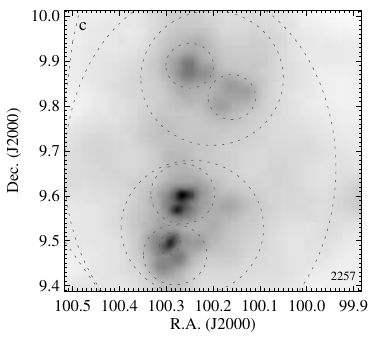}}
\caption{Spatial distribtion for sample ``c'' (center), as in the rightmost panel of Fig.\,\ref{fig:radec_sm}, zoomed in by a factor of four (a 37.5\,arcmin square). We here employed an adaptive kernel smoothing (see text).
}
\label{fig:radec_ctr}
\end{figure}

The map in Fig.\,\ref{fig:radec_ctr} shows the center of the cluster, roughly
within the halo, using sample ``c''. In this case the binned density map was
smoothed using the adaptive kernel smoothing algorithm of
\citet{ebe06}\footnote{With minimum-maximum signal-to-noise ratio (S/N) of
4.0-5.0}. In addition to the structures already identified by \citet{sun08} we
tentatively identify three new substructures: two within the ``S Mon'' region
and one within the Cone(C) region. The first two, named ``S Mon(C)'' and ``S
Mon(ref)'', correspond to two moderate but observable density enhancements;
subregion S\,Mon(C), where C stands for core, is centered on the bright S\,Mon
star, while subregion S\,Mon(ref) is named after the reflection nebula which is
prominent at its location in optical images. The third new subregion,
Cone(C-IR), lies within Cone(C) and corresponds to a compact member overdensity,
which is very prominent in nIR and mIR images including a large fraction of
young and embedded objects (also called IRS1). 

In addition to these regions, close inspection of the density maps in
Fig.\,\ref{fig:radec_sm} may also reveal hints of two filamentary structures: a
vertical one close to the northern edge of our field, at R.A.$\sim$100.3\,deg,
which corresponds to the G02.3+2.5 region, and another one departing at a
$\sim$45 degree angle from the western edge of the ``halo''
(Dec.$\sim$9.5\,deg.) and reaching the edge on the extended halo. We do not
further discuss these possible filaments.

In the following, in order to investigate how SF proceeded in the
region, we characterize the properties of the stellar populations that lie, in
projection, within all of these regions. Field-object contamination, which
varies from region to region, can adversely affect our results. Adopting the
average sky densities from Sect.\,\ref{sect:fied_cont}, it is however small
for all but the outermost subregions. Table\,\ref{tab:subpopcnt} lists,
separately for samples ``b'', ``c-wide'', and ``c'', and for each subregion, the
number of candidate members within the region, the expected net number of
members after subtracting the estimated contamination, and the fraction of
contaminants. For the subregions that contain nested subregions, we provide,
separately, figures referring to the areas that exclude the inner subregions,
and to the whole area within their respective perimeters (indicated by an
asterisk following the region name). The indentations of the region names in the
first column of the table indicate their spatial hierarchy: regions with a given
indentation level contain those listed below it with a higher indentation level.

\newcolumntype{x}{>{\global\let\currentrowstyle\relax}}
\newcolumntype{^}{>{\currentrowstyle}}

\begin{table*}[!th!]
  \begin{center}
  \caption{Population of subregions and expected field-object contamination \label{tab:subpopcnt}}
  \begin{tabular}{xl^r^r^r^r^r^r^r^r^r}
  \hline\hline
  Region    & \multicolumn{3}{c}{Sample ``b''}  & \multicolumn{3}{c}{Sample ``c-wide''} & \multicolumn{3}{c}{Sample ``c''} \\
  \cmidrule(lr){2-4}\cmidrule(lr){5-7}\cmidrule(lr){8-10}
              &  tot &  net & cnt\% &     tot &  net & cnt\% &  tot &  net &cnt\% 	\\            
  \hline
  Field*                           &   1971 &   1859 &  5.7  &   1828 &   1641 & 10.2  &   2257 &   2076 &  8.0 \\
 .  Field                          &    316 &    221 & 30.0  &    449 &    291 & 35.1  &    463 &    310 & 32.9 \\
 .  Ex.Halo*                       &   1655 &   1638 &  1.1  &   1379 &   1350 &  2.1  &   1794 &   1766 &  1.6 \\
 . .  Ex.Halo                      &    322 &    311 &  3.4  &    340 &    322 &  5.4  &    377 &    359 &  4.7 \\
 . .  Halo*                        &   1333 &   1327 &  0.5  &   1039 &   1028 &  1.0  &   1417 &   1407 &  0.7 \\
 . . .  Halo                       &    345 &    341 &  1.1  &    340 &    333 &  1.9  &    387 &    381 &  1.6 \\
 . . .  S Mon*                     &    378 &    377 &  0.3  &    333 &    331 &  0.6  &    406 &    404 &  0.5 \\
 . . . .  S Mon                    &    200 &    199 &  0.5  &    181 &    179 &  0.9  &    219 &    217 &  0.7 \\
 . . . .  S Mon(ref)*              &     77 &     77 &  0.2  &     62 &     62 &  0.4  &     80 &     80 &  0.3 \\
 . . . .  S Mon(C)*                &    101 &    101 &  0.1  &     90 &     90 &  0.3  &    107 &    107 &  0.2 \\
 . . .  Cone*                      &    610 &    609 &  0.2  &    366 &    364 &  0.6  &    624 &    622 &  0.3 \\
 . . . .  Cone                     &    165 &    164 &  0.5  &    130 &    129 &  1.0  &    175 &    174 &  0.7 \\
 . . . .  Cone(C)*                 &    203 &    203 &  0.1  &    121 &    121 &  0.4  &    205 &    205 &  0.2 \\
 . . . . .  Cone(C)                &    149 &    149 &  0.2  &    107 &    107 &  0.4  &    150 &    150 &  0.2 \\
 . . . . .  Cone(C-IR)*            &     54 &     54 &  0.1  &     14 &     14 &  0.3  &     55 &     55 &  0.1 \\
 . . . .  Spokes*                  &    242 &    242 &  0.1  &    115 &    115 &  0.4  &    244 &    244 &  0.2 \\
    \hline
\end{tabular}
\tablefoot{
For samples ``b'', ``c-wide'', and ``c'', and for each subregion, the table provides: the total number of candidates members (tot), the contamination-subtracted number of members (net, see text), and the contamination fraction (cnt\%).
}
\end{center}
\end{table*}

Contamination is clearly an issue mainly for the ``Field'' subregion (excluding
the Extended Halo) where estimates range between $\sim$30\% and $\sim$35\%,
depending on sample. With these levels of contaminations, moreover subject to
uncertainties, characterizing this population is difficult and more data, such
as future GAIA releases, might be needed to determine to what degree the cluster
extends in these outer regions. On the other hand, the population of the other
regions appear to be relatively unpolluted. Even for the Extended Halo$^*$, our
estimates for the contamination are only 1-2\%, rising to 3-5\% when excluding
the inner Halo$^*$ region. The small ``core'' regions (Cone(C)$^*$, Spokes$^*$,
S\,Mon(C)$^*$, and S\,Mon(ref)$^*$) all have negligible contaminations. This
analysis also supports the use of sample ``c'' for the investigation of the
mean (or median) properties of each subregion.

\section{Discussion}
\label{sect:discussion}

We now proceed to exploit our membership determinations, described above, for
our science goals. Our member samples are significantly richer-more complete
with respect to previously available ones. We may, for example, compare our
sample ``c'', comprising 2257 stars, with that assembled by the CSI project
\citep{cod14a}. That sample, based on a early version of our current catalog,
listed 1444 ``very likely'' members. Of these, 322 stars are not included in our
sample ``c''\footnote{186 do not fall in any of our member loci with 3$\sigma$
confidence and are not unique counterparts of X-ray sources, while the remaining
136 are rejected because they are suspected field objects.}, which, however,
includes 1135 new candidates. Overall, we thus almost doubled the known
population of NGC\,2264.

Before focusing on our study of the SF process in the region and on
the early stellar evolution, we make use of our selected member sample to
address the usefulness of the X-ray detection and optical variability as proxies
of stellar youth and, thus, as membership indicators for SFRs. 

\subsection{X-ray detection and optical variability}
\label{sect:xvarsel}

We can exploit our final member selection, sample ``c'', to investigate the
effectiveness of individual membership criteria. We here address X-ray detection,
which is well-established, and ``strong'' optical variability, based on CoRoT
lightcurves (see Sect.\,\ref{sect:corotvar} and Fig.\,\ref{fig:corot}), for each
of our subregions separately\footnote{We exclude from the discussion the
``Field'' region, whose area is not significantly covered by either the X-ray
and CoRoT observations}. Selection efficiency and contamination are both
estimated for each of the two criteria, the former being defined as the fraction
of sample ``c'' members selected by the criterion, and the latter as the
fraction of selected candidate members that are not included in sample ``c''
(that is of those that were rejected as nonmember).

For these statistics, in order to take into account the limited sensitivity of
our X-ray and CoRoT observations, we restrict all the involved samples to stars
brighter than a given limiting optical magnitude. From the G versus BP-RP
diagrams of counterparts to X-ray sources (Fig.\,\ref{fig:GAIAdia}) we set this
limit for X-ray detected stars to G$<$19. We note, however, that the depth of
the X-ray observations varies considerably across the field. For the
variability-based membership we adopt G$<$17 as the completeness limit, from a
similar plot for stars with available CoRoT lightcurves (not shown). We caution
that the X-ray detections were actually used to define our reference member list
(sample ``c''); assuming that our reference sample is reasonably complete, down
to the adopted limiting magnitude, this should not overly bias the statistics.
On the other hand, the CoRoT data were not used to define sample ``c''. However,
the targets observed by CoRoT were selected in advance and, in the central part
of the field, roughly within the Halo region, are biased toward known
NGC\,2264 members. This may well bias the estimated contamination fractions in
these regions, but should not affect the selection efficiency.

Table\,\ref{tab:subregions} lists, for each of our subregions (including the
whole considered area, that is the Field$^*$): the number of candidate members in
the samples (X$_N$ and CoRoT$_N$), the selection efficiencies (X$_{sel.f}$ and
CoRoT$_{sel.f}$) and the contamination fractions (X$_{rej.f}$ and
CoRoT$_{rej.f}$), both for X-ray detection and for the variability analysis
based on CoRoT data. For X-ray detection the selection efficiency ranges from
81-92\% in the dense Cone, Cone$^*$, Cone(C), and Spokes regions, which are
deeply imaged in X-rays, to $\sim$50\% in the Halo and Extended Halo regions,
which are instead mostly or exclusively covered by the relatively shallow
{\em XMM-Newton} observations (cf. Fig.\,\ref{fig:FOV}). Indeed,
Table\,\ref{tab:subregions} also shows the result of limiting the sample to
G$<$17, instead of G$<$19, yielding higher and more consistent selection
efficiencies. Contamination from objects rejected as nonmembers by our
selection procedure is typically $<$10\% in the densest regions, but reaches as
high as $\sim$32\% in the low-density Extended Halo.

From the selection efficiencies listed in Table \ref{tab:subregions} we infer
that member selection based on optical variability (\S\,\ref{sect:corotvar}) is
as effective as X-ray detection. As for contamination fractions, because of the
mentioned target selection bias, the only reliable figure is probably that for
the Extended Halo region, in which no members were known at the time of the
CoRoT target selection. Although relatively high, 42\%, it is comparable with
that for X-ray detection in the same region (32\%). Even considering that these
results are based on a specific analysis of CoROT lightcurves, having specific
and favorable characteristics in terms of cadence and duration, these estimates
suggest that optical variability may prove a powerful method for selecting young
stars in SFRs, and may be particularly useful when applied to forthcoming large
datasets, such as those from GAIA and Rubin-LSST \citep[see, for example,][]{bon18}.

\subsection{Physical properties}

We now investigate the properties of the NGC\,2264 population, both globally and
for each of the subregions we have defined. As described in
\S\,\ref{sect:ResSpatialAndCont}, we have tentatively identified four new
substructures in addition to those of \citet{sun09}: the Extended Halo,
significantly enlarging the known extension of the cluster in the plane of the
sky ($\times$2.7 the area of the Halo), two substructures within the S\,Mon
region (the S\,Mon ``core'', S\,Mon(C)$^*$, around the bright O-type star, and
S\,Mon\,(ref)$^*$, an apparently looser subclustering, southwest of S\,Mon,
corresponding to a reflection nebula in the optical images), and the
Core(C-IR)$^*$ region within Cone(C), corresponding to a clear compact density
enhancement, also prominent in nIR and mIR images. In the following we mostly
ignore our ``Field'' subregion which, as discussed in
\S\,\ref{sect:ResSpatialAndCont}, may or may not contain a significant
population of NGC\,2264 members. We begin assessing the spatial and kinematic
properties of these structures and then discuss stellar masses, ages, and
circumstellar disk frequencies. We will often refer to
Table\,\ref{tab:subregions}, which lists several average or median properties of
stars in each subregion.

\subsubsection{Spatial distribution and densities}

\begin{figure}[h]
\centering
\resizebox{0.45\textwidth}{!}{\includegraphics{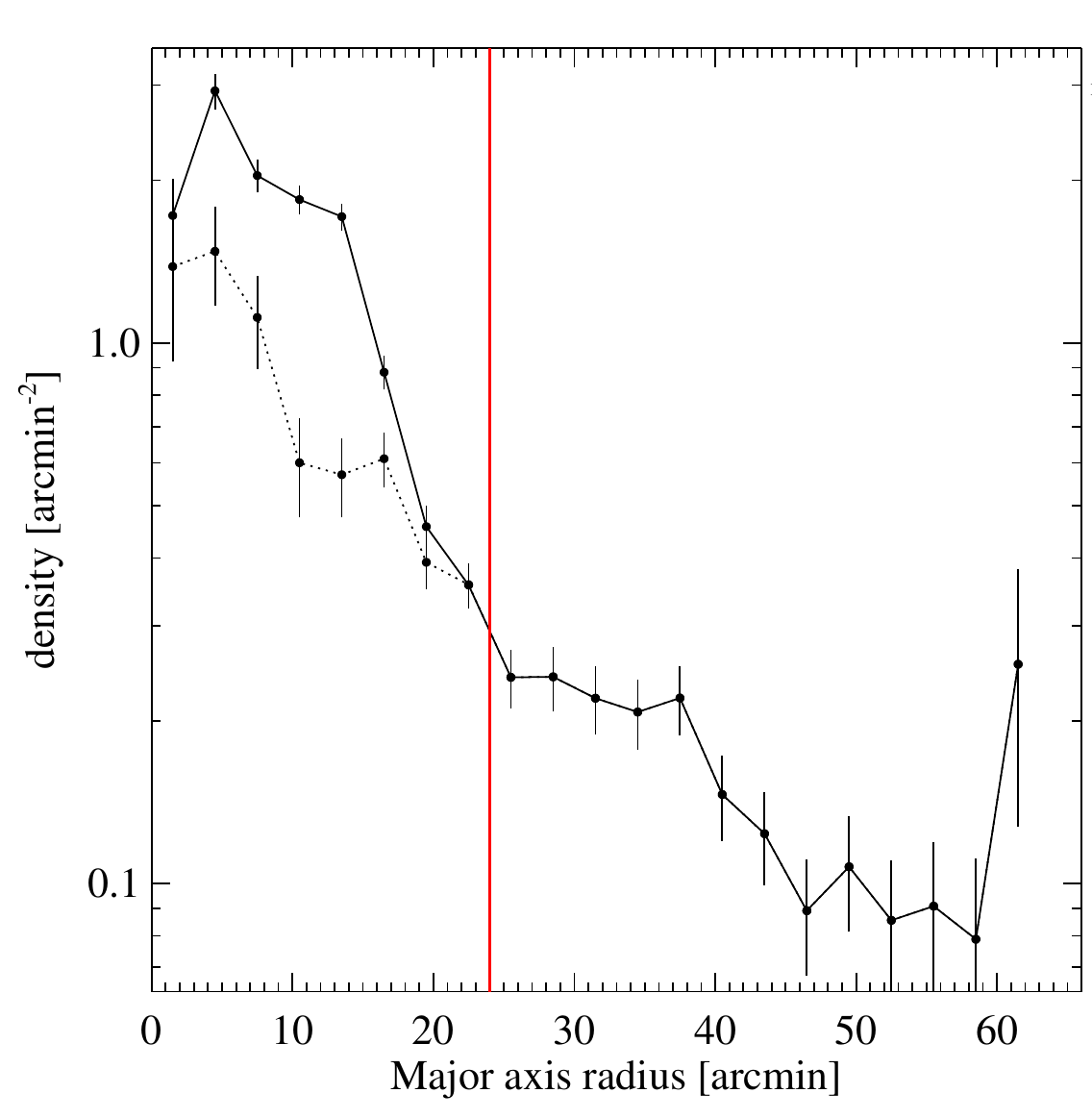}}
\caption{Density profile of sample ``c'' members within the Extended Halo$^*$
subregion. Densities plotted as a solid line are computed in the intersection
between the Extended Halo$^*$ region and elliptical annuli, concentric with those
used in Fig.\,\ref{fig:field_cont}. The dotted line is the result of excluding
the dense S\,Mon$^*$ and Cone$^*$ region. The vertical red line indicates the boundary
of the Halo region.}
\label{fig:densprof}
\end{figure}

Our substructures span a wide range of average stellar densities (cf.
Table\,\ref{tab:subregions}), a factor of 28(55) between the Extended Halo and
the Spokes$^*$ cluster (Cone(C-IR)$^*$ region). Figure\,\ref{fig:densprof} shows
the density profile within the Extended Halo$^*$ region, computed in elliptical
annuli concentric with the Halo perimeter (cf. \S\,\ref{sect:ResSpatialAndCont}
and Fig. \ref{fig:field_cont}). Two profiles are shown, both including and
excluding the denser S\,Mon$^*$ and Cone$^*$ regions. A break in the density
profiles, at a radius similar to the boundary between the Halo and the Extended
Halo, possibly suggests that the two structures might indeed be distinct, in
spite of the fact that most of the other average stellar properties, discussed
below, appear to be compatible. Note that the density of the plateau at radii
larger than 45\arcmin, at $\sim$0.08 stars/arcmin$^2$, is $\sim$10 times the
density of the field-object contaminants, as estimated from
Fig.\,\ref{fig:field_cont}.

The external parts of the two main subclusters, the S\,Mon and the Cone
regions, have similar densities, $\sim$1 star/arcmin$^2$, but the two cores
within the Cone region are slightly denser than those in the S\,Mon region
(4.0-4.8 versus 2.8-3.8 stars/arcmin). The densest structures are indeed the
Cone(C-IR)$^*$ and Spokes$^*$ subcluster(s) dominated by an embedded and
presumably very young population (9.3 and 4.8 stars/arcmin$^2$, respectively on
average, with a central peak of $\sim$30 stars/arcminn$^2$ in both cases).

\subsubsection{Distances}

We estimate the distance to the NGC\,2264 region from the median of the
individual GAIA parallaxes of 1337 members (in our sample ``c'') within the
Ex.Halo$^*$ region. We choose not to use the other candidate members, in the
``Field'' region, because of the potentially significant contamination from
nonmembers. Our best median distance estimate is thus of 724$\pm$8\,pc. We also
estimate a distance of 722$\pm$2\,pc, excluding 211 ``outliers'', with
parallaxes outside of the 1.1-1.7\,mas range\footnote{We also considered
adopting an uncertainty weighed median, which yield 722$\pm$8\,pc, or
721$\pm$2\,pc excluding outliers. However, this implicitly assumes that all
stars are at the same distance.}.

\begin{figure*}[!t]
  \centering
  \resizebox{0.245\textwidth}{!}{\includegraphics{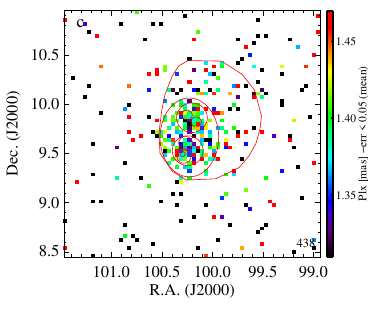}}
  \resizebox{0.245\textwidth}{!}{\includegraphics{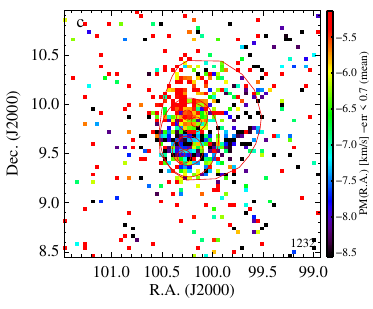}}
  \resizebox{0.245\textwidth}{!}{\includegraphics{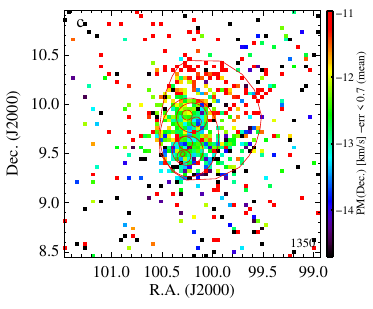}}
  \resizebox{0.245\textwidth}{!}{\includegraphics{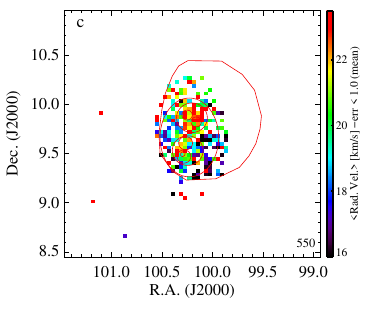}}
  \resizebox{0.245\textwidth}{!}{\includegraphics{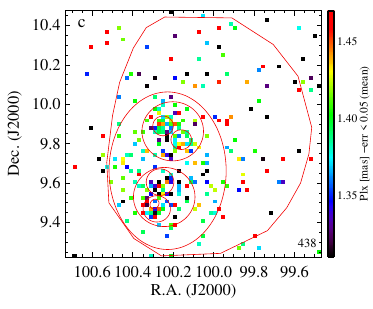}}
  \resizebox{0.245\textwidth}{!}{\includegraphics{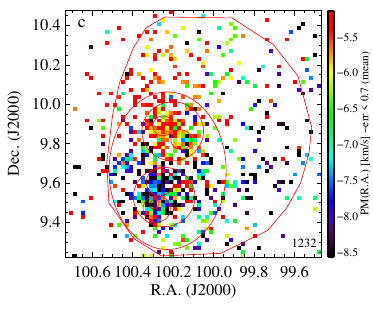}}
  \resizebox{0.245\textwidth}{!}{\includegraphics{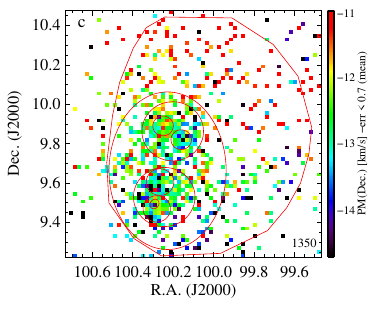}}
  \resizebox{0.245\textwidth}{!}{\includegraphics{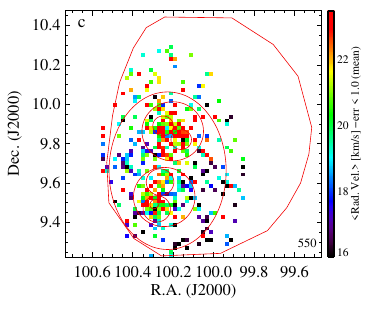}}
 \caption{Spatial distributions for sample ``c'', color-coded according to four
  different variables. The top row refers to our full FoV, while the bottom row shows the central area in better detail. Within each row, the panels show, from left to right: parallax, PM$_{R.A.}$, PM$_{Dec.}$, and
  radial velocity. Color are assigned using the average value of the variable
  for all stars within each pixel and following the scale on the right-hand side
  of each panel. Most pixels on the outskirts contain a single star. The
  numbers on the bottom left indicate the number of stars used for the panel.
  }
  \label{fig:cc_maps_avg}
\end{figure*}

In order to investigate possible differences in distance across the region,
Fig.\,\ref{fig:cc_maps_avg} shows a binned color-coded map of the region, with
each pixel indicating the average parallax of the stars within that pixel (most
pixels in the outer sparse regions contain only one star). Because of the
relatively large uncertainties on parallaxes, we limit the sample to stars for
which these uncertainties are smaller than 0.05\,mas. No outstanding
pattern is visible across the whole region. However, the bulk of the stars in
the Halo$^*$ region share a similar parallax ($\sim$1.4\,mas, in green), while
the stars in the Field region seem to be either farther away (blue) or closer
(red). Within the Halo we also note that the Cone$^*$ and the S\,Mon$^*$ regions
contain several clustered blue and black areas, with the main concentration
corresponding to the embedded Spokes$^*$ cluster.

\begin{figure}[h]
  \centering
  \resizebox{0.45\textwidth}{!}{\includegraphics{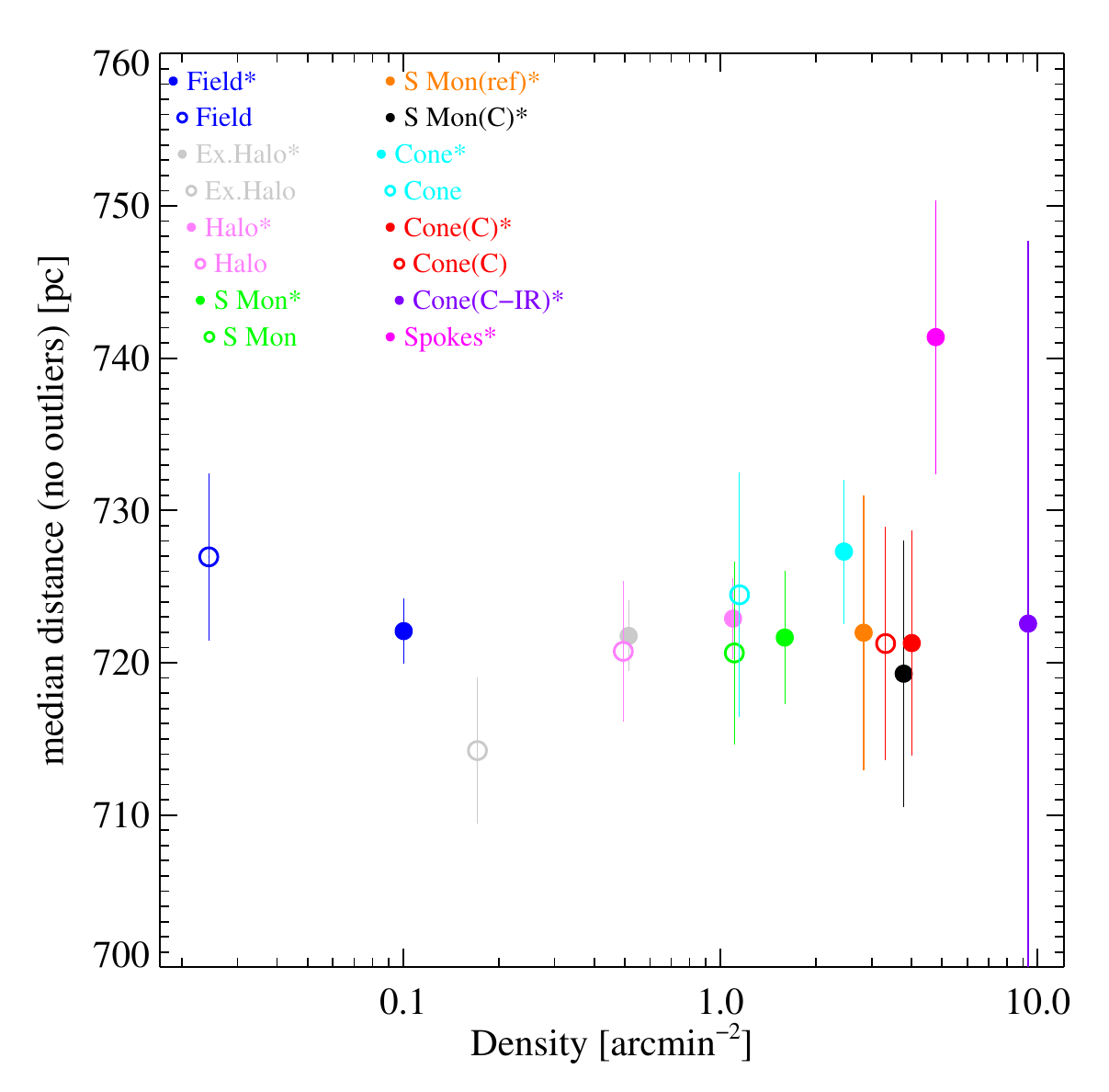}}
  \caption{Distance vs. average stellar density for all the defined subregions. Distances are computed as the median of all stars with GAIA parallaxes in each subregion, excluding outliers (see text). Error bars refer to uncertainties on the median values. Filled symbols refer to the whole area within a region perimeter, while empty symbols to ``haloes'', i.e. excluding nested regions, if present. The color coding is as in the legend.}
  \label{fig:dist_vs_dens}
\end{figure}

Figure\,\ref{fig:dist_vs_dens} shows the median distance for all the defined
subregions, here computed excluding outliers (parallaxes outside of the 1.1-1.7\,mas
range) as a function of average stellar density. The distances of most of the
subregions are compatible with the adopted NGC\,2264 distance, that is that of the
Ex.Halo$^*$ region, indicated in the figure by the filled gray circle. The most
notable exception is the Spokes$^*$ regions, which seems to be $\sim$20\,pc
farther away than the rest of the cluster. When considering the statistical
uncertainties on the medians, the discrepancy between the Spokes$^*$ and
Ex.Halo$^*$ distances is significant at the 2.1$\sigma$ level, rising to
2.7$\sigma$ when comparing with the Ex.Halo region (the outer population)
which might be at a slightly closer distance.

Similar conclusions can be drawn comparing the distributions of the distances of individual stars
within each subregion, again excluding distance outliers, using two-sided
Kolmogorov-Smirnov (KS) tests. The probability that the distribution of
distances for stars in the Spokes$^*$ and Ex.Halo$^*$ regions are drawn from the
same parent population is 0.52\%, which is reduced to 0.02\% when comparing the
Spokes$^*$ and Ex.Halo (outer population only) regions. Similar but slightly
less significant null probabilities, 1.4\% and 0.03\%, are obtained when also
including parallax outliers, but excluding values with uncertainties $>0.3$
mas.
 
We also note that these results are obtained in spite of the small fraction of
available parallaxes among the members in the Spokes$^*$ region: 33\% and 36\%
for the no-outlier and uncertainty-restricted samples, respectively,
significantly lower than those for the corresponding samples in the Ex.Halo$^*$
region, which are 63\% and 64\%, respectively. This is easily understandable since we
expect that the most embedded stars, which are optically faint, are less likely
to have measured and good-quality GAIA parallaxes. The true median distance of the
Spokes$^*$ region might thus be even larger than estimated with the available
parallaxes.
  
We conclude that the Spokes$^*$ cluster is very likely behind the main NGC\,2264
population. Given the uncertainties, the exact depth separation is only loosely
constrained. It is, however, of the same order as the extent of the star-forming
molecular cloud in the plane of the sky, as traced by the population we have
selected: $\sim$1$\deg$ or $\sim$12.5\,pc.

\subsubsection{Kinematic properties}

The second, third, and fourth panels of Fig.\,\ref{fig:cc_maps_avg} show binned
spatial maps of members, color coded according to the mean R.A. and Dec.
proper motions, and radial velocity, respectively. The proper motion scale was
converted to km/s, assuming a distance of 722\,pc. Patterns and trends are
clearly observable, but obtaining a clear picture of the structures and the
nature of their motion is not straightforward. Velocities do not appear to be
strictly associated with the subregions we have defined on the basis of the
spatial distributions. We tentatively recognize signs of expansion in all three
directions, more markedly in the north-south direction, consistent
with the overall elongation of the cluster. It also seems that the Halo region
is expanding toward the viewer with respect to the denser inner subregions
(Cone$^*$ and S\,Mon$^*$) at 2-3\,km/s.

Rotation is also a possibility, mostly in the plane of the sky. On average, the
southern region (Cone$^*$) moves toward the west at $\sim$0.5\,mas/yr (1.7 km/s)
relative to the northern S\,Mon$^*$ region. This effect can be significant, as
it translates to 25 arcmin (5.5pc) in 3Myr (the approximate age of the cluster).
The region might thus have rotated significantly since collapse started.

\begin{figure*}[!t]
  \centering
  \resizebox{\textwidth}{!}{\includegraphics{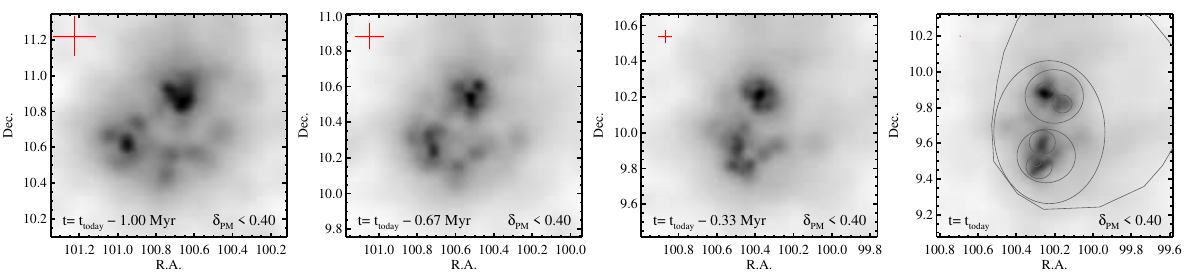}}
  \caption{Spatial densities predicted at three past epochs: 1, 0.67, and
  0.33\,Myr (from left to right) in the past, plus the present one (on the
  right). As indicated in the panels, only stars with PM uncertainties $<$0.4\,mas/yr
  are used and the maximum 1$\sigma$ uncertainties for the reprojected position
  are indicated by the red cross in the upper-left corner. The density maps are
  smoothed with an adaptive kernel as in Fig.\ref{fig:radec_ctr}. 
  }
\label{fig:traceback}
\end{figure*}

Figure \ref{fig:traceback} shows the evolution of stellar density in the last
million years as derived by simply back-tracing the coordinates of members with
measured proper motions. This extrapolation assumes rectilinear motions and
no intervening SF. Only stars with PM uncertainties $<$0.4 mas/yr
are used and, in the four panels representing epochs from 1\,Myr  ago to the
present time, the red cross in the upper left corner indicate the maximum
$1\sigma$ uncertainty in individual stellar positions. We observe the overall
counter-clockwise rotation discussed above. Probably more interestingly, we see
clear signs of collapse: two structures, at least one with a filament-like shape,
appear to fall toward each other in the southern regions of the cluster, forming
the present day substructures we named Cone(C) and Stokes. Both of these regions
contain a rich embedded and extremely young population, very likely much younger
than 1\,Myr. We argue that the compression, or density increase, due to the collapse
of these two structures is the cause of this recent formation outbursts.

This scenario is somewhat similar to that inferred by \citet{mon19b} for
G02.3+2.5, $\sim$0.2-1.0\,$\deg$ north of S\,Mon, from the radial motions of its
gas and dust filaments. Although the region is for the most part included in our
field, we do not have evidence of this collision from stellar motions on the
plane of the sky. It might be more evident in stellar radial velocities, which
are not available in this region or, alternatively, we might simply have too few
stars with proper motions in this region to notice the effect, especially since
many young stars may be embedded and lack good GAIA astrometry.

\subsubsection{Masses and ages}
\label{sect:massesages}

We estimated masses and ages for our candidate members using the MIST (v. 1.2)
evolutionary tracks \citep{cho16,dot16}. Four different sets of masses and ages
were derived by interpolating the evolutionary tracks for rotating
($v/v_{crit}=0.4$), solar abundance stars, in different planes: photometric CMDs
from the IPHAS, Pan-STARRS, and GAIA surveys (i versus r-i, for the first two
surveys and G versus BP-RP for GAIA), and the fundamental plane (L$_{bol}$ versus
T$_{eff}$). For this latter, spectral types (available for 433 members in sample
``c'') were used to estimate effective temperatures, bolometric correction, and,
together with photometry, extinction values for each star\footnote{Temperature
scales from \citet{ken95} and \citet{luh99b} (intermediate scale) for spectral
types earlier and later than M0, respectively. Bolometric correction, BC$_I$,
from \citet{ken95}. For extinction: intrinsic R-I$_c$ colors from \citet{ken95},
A$_V$=4.46 E(R-I$_c$), A$_I$=A$_V$/1.63 }. These individual extinctions values,
were also used to derive mean values, which we adopted to de-redden the stars in
the photometric planes before interpolating the tracks.

The four sets of masses and ages were then compared among each other, obtaining
reasonable agreements: the 1-$\sigma$ logarithmic dispersion between all the
different pairs of estimates range from 0.06 to 0.14\,dex for masses and between
0.18 and 0.40\,dex for ages. We finally merged the four estimates, assigning the
`best' available values to each star, with the following precedence with respect
to the origin of the estimates: the L$_{bol}$ versus T$_{eff}$
spectro-photometric plane (with individual extinction corrections), Pan-STARRS,
IPHAS, and, finally, GAIA photometry. While we expect these estimates to be
reasonable for the majority of NGC\,2264 members, we warn that, for unresolved
binaries and for heavily absorbed or embedded stars without spectral types, and
therefore with no individual extinction estimates, estimated masses and ages
might be rather biased. An additional small bias might also be introduced by
variation in distances, for example making the embedded and more distant stars
in the Spokes region look fainter and, therefore, older than they really are.    
  
\begin{figure*}[h]
  \centering
  \resizebox{0.45\textwidth}{!}{\includegraphics{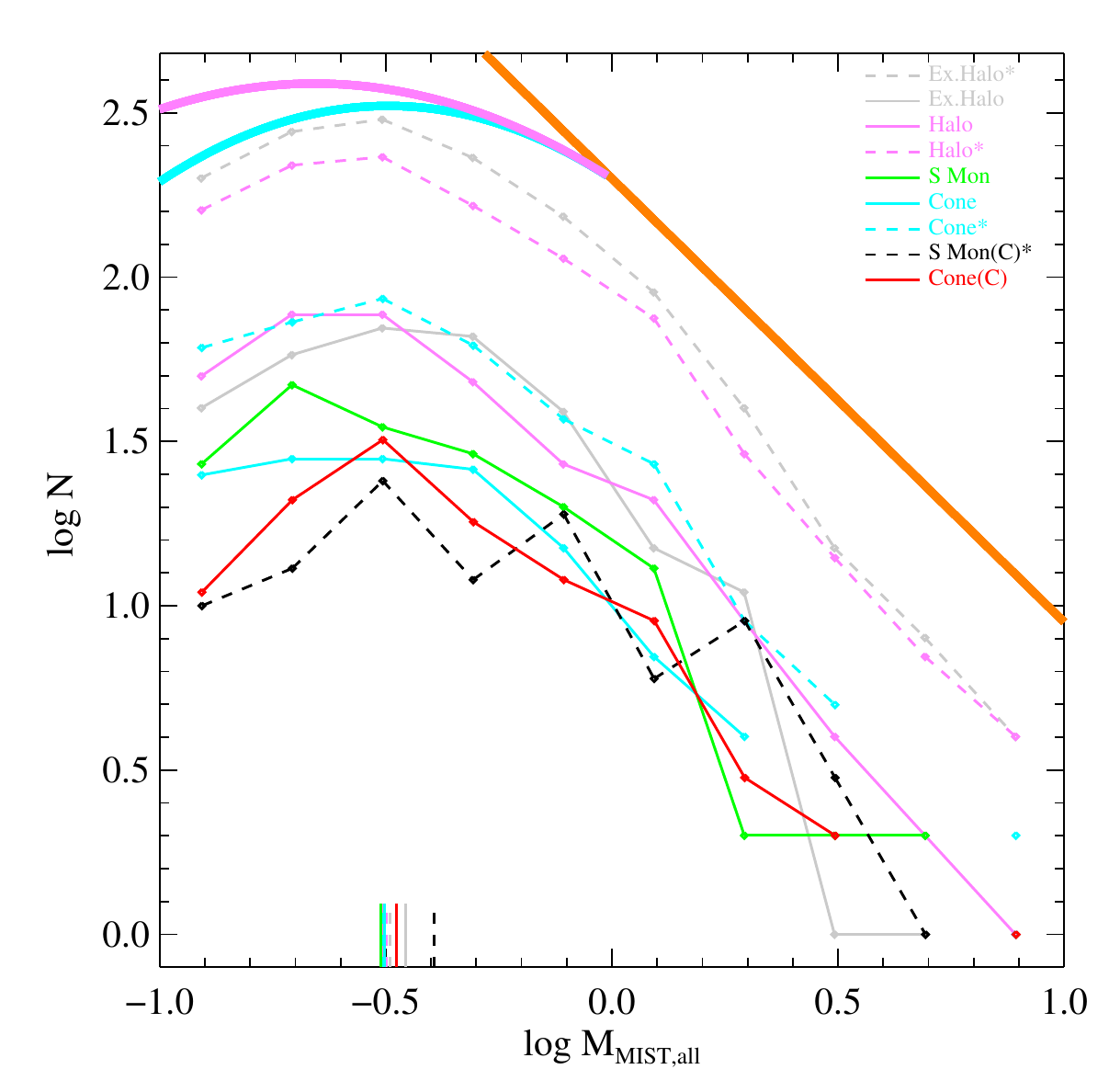}}
  \resizebox{0.45\textwidth}{!}{\includegraphics{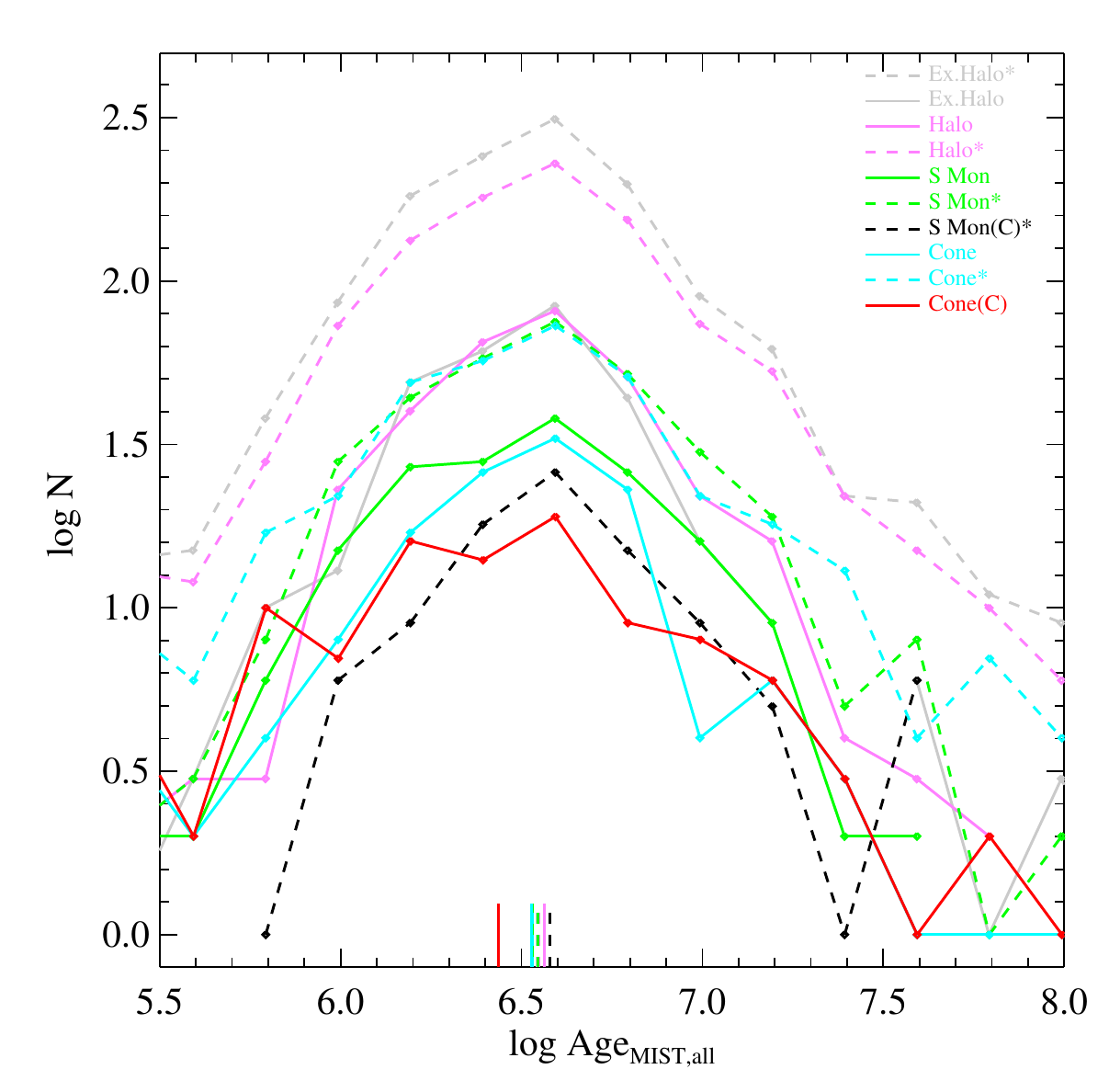}}
  \caption{Logarithmic mass and age distributions (left and right panels,
  respectively) for selected subregions, color coded as in the caption in the
  upper right corner. For regions with nested subregions, dashed lines refer to
  the whole area within their perimeters, and solid lines to the `haloes'. Thick
  solid lines in the left panel show the Salpeter's Initial Mass Function (IMFs, \citealt{sal55}, slope
  -1.35) in orange, and the \citet{cha03} system and single stars IMF, for
  M<M$_\odot$, in cyan and magenta, respectively. The short vertical lines near
  the x-axes, at log\,M$\sim$-0.5 and log\,Age$\sim$6.5, indicate the median
  values of each samples.}
  \label{fig:MAdist}
\end{figure*}

Figure \ref{fig:MAdist} shows the mass and age distributions for the whole sample
``c'' (within the Extended Halo) and for some of the subregions. No striking
differences are seen for masses. However, KS tests suggest, with a $\sim$99\%
confidence, that the S\,Mon(C) subregion might harbour more massive stars when
compared to the Halo, the S\,Mon, and the Cone$^*$ subregions. This is in spite
of the fact that the most massive star in the region, the O7 star S\,Mon, is not
included in the mass distribution.

The age distributions of most subregions are statistically compatible, when
comparing them using KS tests. The exception seems to be Cone(C), which does
appear to be somewhat younger than other subregions, with null probabilities
$<$1\% when compared to the Ex. Halo$^*$, the Ex.Halo, the Halo, and
S\,Mon(C)$^*$ (n.p. =0.9\%, 0.8\%, 0.2\%, and 0.3\%, respectively, when
comparing the distributions of ages plotted in
Fig.\,\ref{fig:MAdist})\footnote{If we consider only spectral-type based ages,
the distribution of ages for the Cone(C) differs, with n.p.$<$1\%, only from
those of the S\,Mon$^*$ and S\,Mon subregions (n.p. = 0.19\% and 0.15\%,
respectively)}. For some of the subregions, notably for the Spokes$^*$ and the
Cone\,(C-IR)$^*$ regions (not shown in Fig.\,\ref{fig:MAdist}), the fraction of
stars with age estimates is small, 41\% and 33\%, respectively, and their median
ages are likely overestimated. Indeed, their embedded and likely young
populations may well be under-represented, and, as noted above, the ages that
are available might be adversely affected by the unaccounted-for larger
extinctions and distances. Moreover, all of our age estimates may be affected by
the presence of accretion disks, through the luminosity excesses and color
shifts produced by accretion spots (in the blue part of the spectrum or
photometric bands), thermal emission by inner disks (in the red bands), and
scattering of stellar emission by the disks. Since, as shown in the next
section, the fraction of stars with accretion disks varies significantly in
different subregions, even the relative age scale might be skewed. We thus
consider our median age estimates with extreme caution.

\subsubsection{Disk and accretion fractions}

We investigate differences in disk and accretion fractions across the SFR, using
both binned spatial maps and focusing on the subregions we have defined. We
define the subsample of members with disks as the stars that fall in one of the
two mIR-based member loci ([3.6]-[5.8] versus [4.5]-[8.0] and W1-W2 versus W2-W3, in
Fig.\,\ref{fig:diagrams_sampc}), with 2$\sigma$ confidence\footnote{For
membership we required a 3$\sigma$ confidence.}. Likewise, we define as
accreting members those that lie, with 2$\sigma$ confidence, in at least one of
the five members loci based on H$\alpha$ emission, as observed either
photometrically or spectroscopically (Fig.\,\ref{fig:GAIAdia} and
\ref{fig:diagrams_sampc}). Note that having independent accretion assessment is
useful, not only to enlarge the sample of stars with appropriate data, but also
to account for the variable nature of accretion.

\begin{figure*}[!h!]
  \centering
  \resizebox{1.0\textwidth}{!}{\includegraphics{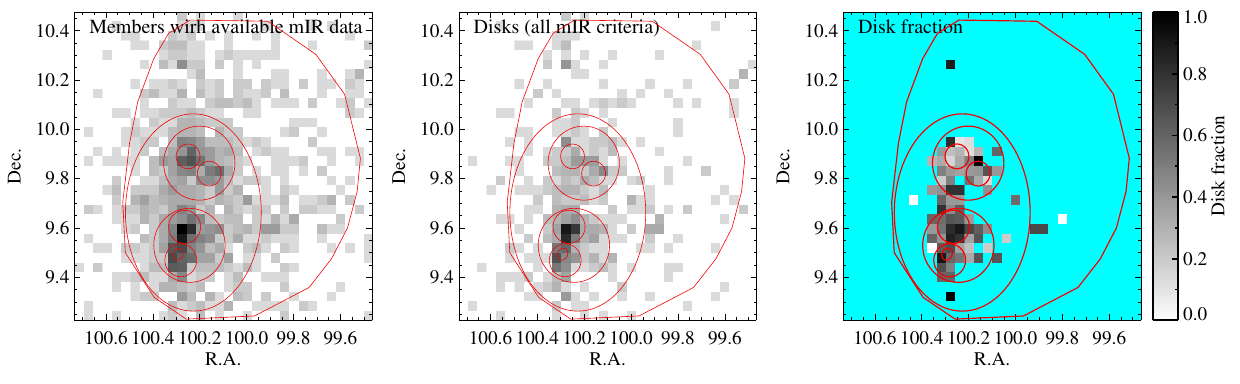}}
  \caption{Binned spatial maps of (from left to right): i) stars placed in
  either of the two mIR based diagrams from which we assess the presence of
  disks, ii) stars with disks, iii) the fraction of stars with disks. In the
  rightmost plot the gray scale follows the color-bar on the right-hand side, and
  the spatial bins in which the denominator of the fraction contained less than
  5 stars are plotted in cyan. 
  }  
\label{fig:mIRexf_map}
\end{figure*}

\begin{figure*}[h]
  \centering
  \resizebox{1.0\textwidth}{!}{\includegraphics{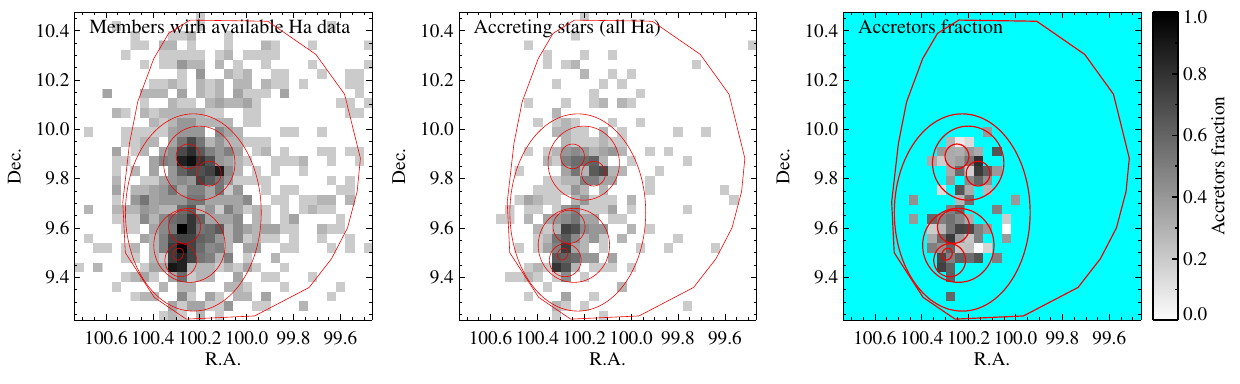}} 
  \caption{Same as Fig.\,\ref{fig:mIRexf_map}, but for the accretion
  indicators. The leftmost panel shows stars placed in any of the five H$_\alpha$ based diagrams, the central one those lying in any of the five accretion loci, and the rightmost one the ratio between the former two maps.}
  \label{fig:Haf_map}
\end{figure*}

Figure\,\ref{fig:mIRexf_map} shows, in the left-hand panel, the binned map of
candidate members with sufficient mIR data to determine disk presence; the
central panel refers to the subset of stars with disks, and the one on the right
shows the fraction of stars with disks in each spatial bin. Although the former
two maps cover the whole region, in order to reduce Poisson uncertainties, we
only show the disk fractions for the spatial bins containing a minimum of 5
members per bin (in the left-most panel). The map, which is therefore only
defined within the dense cores, shows that the Cone region has 
larger disk fractions than the S\,Mon region. Figure\,\ref{fig:Haf_map}
similarly shows maps for accreting stars. Note that the heavily
absorbed or embedded regions, such as the Spokes cluster, are quite clearly
under-sampled by optical observations and that, consequently, their accretion
fraction are likely underestimated.

\begin{figure}[h]
  \centering
  \resizebox{0.45\textwidth}{!}{\includegraphics{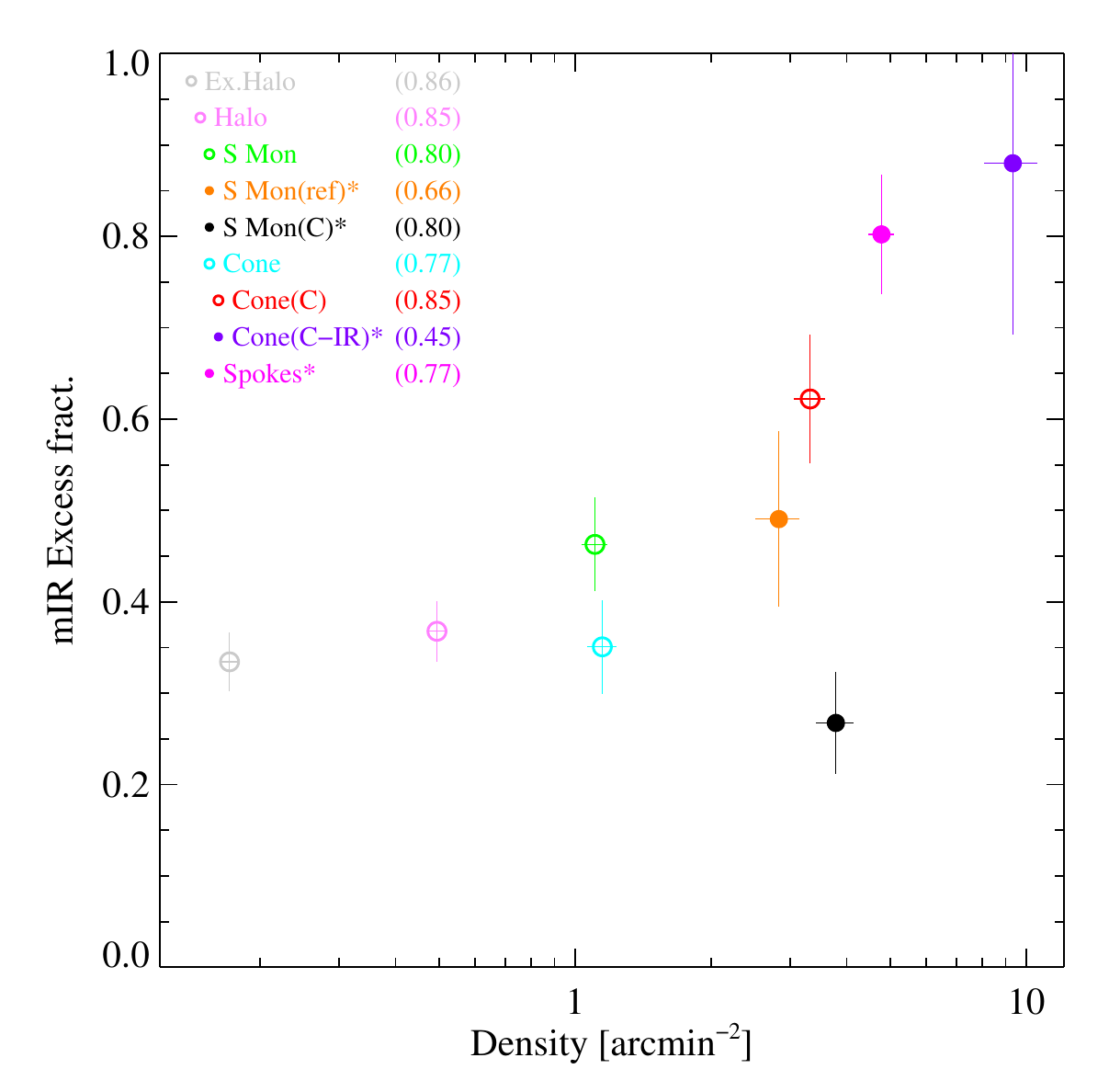}}
  \caption{Mid-IR excess vs. stellar density for independent regions. Symbols and colors as in Fig.\,\ref{fig:dist_vs_dens}. The numbers to the right of the region names are the relative fraction of stars for which the disk indicators are available.}
  \label{fig:AccDiskf_dens} 
\end{figure}

Highly significant differences are seen when comparing the disk fractions of our
subregions: from 80-88\% for the Spokes$^*$ and Cone(C-IR)$^*$ regions to
$\sim$27\% for the S\,Mon(C) region. This is shown in
Fig.\,\ref{fig:AccDiskf_dens}, where we plot the disk fractions versus the mean
stellar densities of the subregions. A correlation is seen, with the possible
outlier of the S\,Mon region, which appears to have a factor of $\sim$2 lower
disk fraction with respect to the general trend. An even clearer correlation is
seen when plotting accretion fractions (instead of disk fraction) versus density
(not shown, data in Table\,\ref{tab:subregions}), again with a noticeable
$\sim$30\% reduction in accretors fraction for the S\,Mon region.

\begin{figure}[h]
  \centering
  \resizebox{0.45\textwidth}{!}{\includegraphics{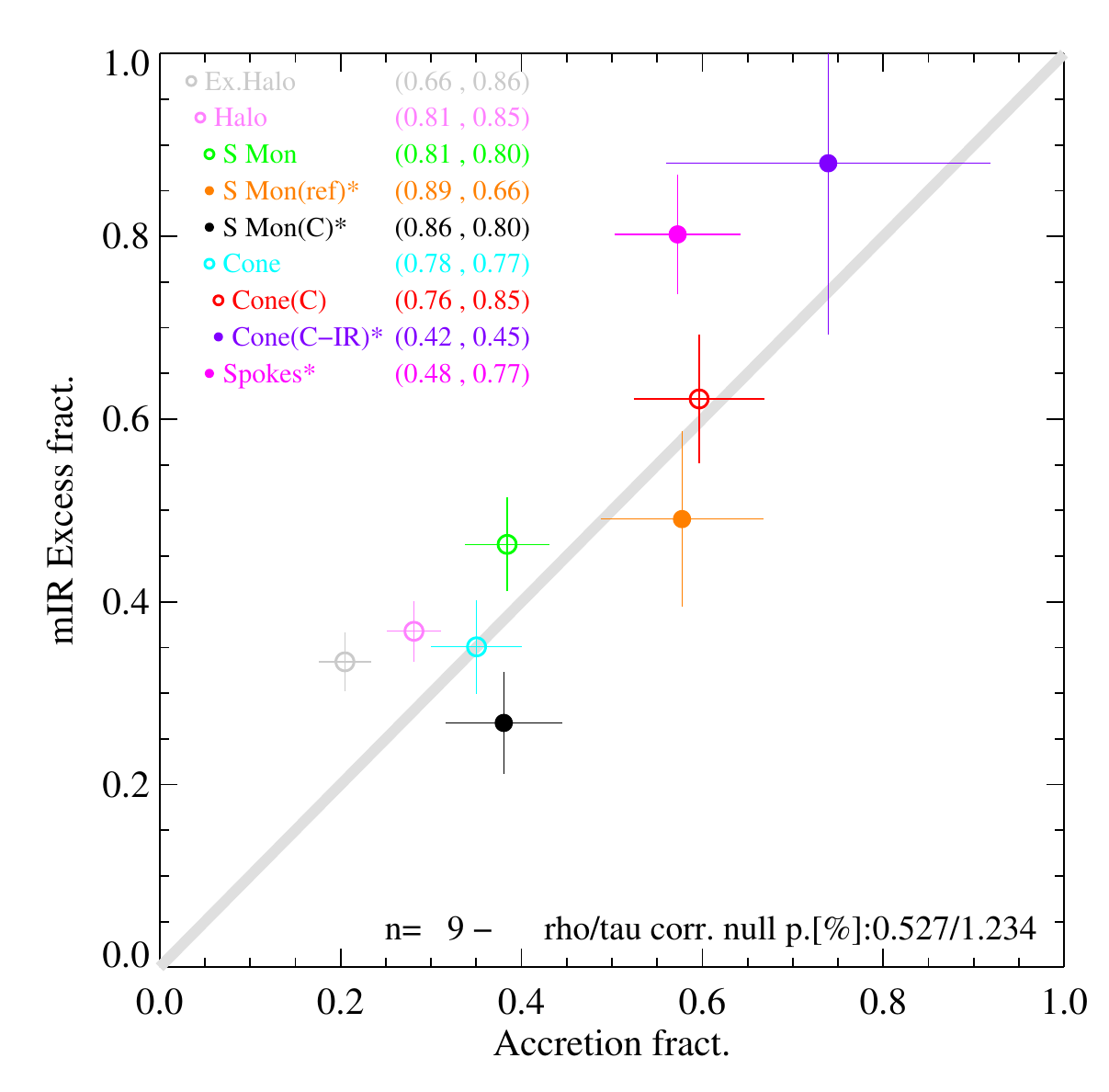}}
  \caption{Disk fractions vs. accretion fractions for independent subregions.
  Symbols and colors as in Fig.\,\ref{fig:dist_vs_dens}. The two numbers beside each region name are the relative fraction of stars for which indicators of accretion (left) and disks (right) are available. The identity line is
  shown in gray and the results of Spearman’s $\rho$ and Kendalls’s $\tau$
  correlation tests are shown in the bottom right.}
  \label{fig:DiskAcc_f}
\end{figure}

Indeed, the accretion fractions estimated from $H\alpha$ emission correlate well
with the disk fractions from the mIR data (Fig.\,\ref{fig:DiskAcc_f}).  
Spearman’s $\rho$ and Kendalls’s $\tau$ tests give 0.5-1.2\% probabilities that the
two fractions are uncorrelated. The Spokes$^*$ region seems to have somewhat
fewer accretors than stars with disks, which  might be explained as a selection
effect, that is by a lower fraction of available H$\alpha$ data among the embedded
(young) stars. Similarly, the deficit of accreting stars in the Ex. Halo might
be attributed to the fact that the outer regions are covered by only one
$H\alpha$ survey (IPHAS) while the central region is covered by up to 5
datasets. When using the same samples, that is stars with both disk and
accretion information, the correlation between the two fractions becomes even
more significant, in spite of the resulting larger uncertainties (null
probabilities: 0.16\% and 0.67\%). Moreover, all accretion and disk fractions
become compatible at the $\sim 1.5\sigma$ level. We can conclude that our data
is insensitive to possible differences between the timescales for the evolution
of accretion and disks.

We tried to correlate disk and accretion fractions with stellar ages but,
unfortunately, as discussed (Sect.\,\ref{sect:massesages}) median ages for the
subregions are affected by large uncertainties, both statistical and systematic.
Disregarding the two heavily embedded regions, for which unaccounted-for
extinctions and selection effects probably dominate, the subregion with the
highest disk and accretion fractions, Cone(C), might indeed be the youngest, while
the S\,Mon(C)$^*$ region, the one with the lowest disk and accretion fractions, might
be the oldest (cf. Fig.\,\ref{fig:MAdist} and KS tests discussed in
Sect.\,\ref{sect:massesages}), as expected but rarely, if ever, observed within
a single star-forming region. However, we do not report this as a result, as we
are concerned that age determinations might be affected by the presence of disk-
and accretion-related excess, indeed spuriously producing such a correlation.

\begin{table*}[!h!] 
\begin{center}
\caption{Properties of the populations within subregions (sample ``c'')\label{tab:subregions}}
\resizebox{1.00\textwidth}{!}{
\begin{tabular}{lrrrrrrrrrrrrrrrrrrrrrr}
\hline\hline
\multicolumn{2}{r}{Field*}&\multicolumn{3}{c}{Ex.Halo*}&\multicolumn{1}{r}{Halo*}&\multicolumn{2}{r}{S Mon*}&\multicolumn{3}{c}{S Mon(ref)*}&\multicolumn{1}{r}{Cone*}&\multicolumn{2}{r}{Cone(C)*}&\multicolumn{3}{c}{Cone(C-IR)*}\\
&\multicolumn{2}{r}{Field}&\multicolumn{3}{c}{Ex.Halo}&\multicolumn{1}{r}{Halo}&\multicolumn{2}{r}{S Mon}&\multicolumn{3}{c}{S Mon(C)*}&\multicolumn{1}{r}{Cone}&\multicolumn{3}{c}{Cone(C)}&\multicolumn{1}{r}{Spokes*}\\
\expandableinput subregions_med.tex
\hline 
\end{tabular}
}
\tablefoot{
For the candidate members in each subregion we list, in order: the total number, and those for stars with G$<$17 and G$<$19; the mean star sky-density; the number of stars with G$<$19 unanbiguosly detected in X-rays; the selection efficiency and the rejection fraction for this latter sample; the same three quantities for G$<$17; the number stars with G$<$17 unanbiguosly observed with CoRoT, and the relative selection efficiency and rejection fraction; the fraction of stars with mIR information useful to detect mIR excesses; the fraction of this latter sample showing mIR excesses and relative statistical uncertainty; the fraction of stars with H$\alpha$ data; the fraction of this latter sample with H$\alpha$ detected in emission and relative statistical uncertainty; the fraciton of stars with distance estimates, the median distance, and the relative uncertainty; the fracition of stars with distance estimates excluding outlier values, the median distance for this sample and the relative uncertainty. The fraction of stars with P.M. estimates; the median P.M. in the R.A. and Dec. coordinates, followed by the relative uncertainties; the fraction of stars with R.V. estimates, the median R.V. and relative uncertainty.
}
\end{center}
\end{table*}

\section{Summary and conclusions}
\label{sect:conclusions}

Following the recognition that a significant number of stars far from the known
NGC\,2264 members show strong optical variability, similarly to the known
members, we embarked on a novel investigation of the structure of this well
studied star-forming region. First, we have obtained new X-ray imaging data,
collected with {\em XMM-Newton}, significantly enlarging the area in which X-ray
sources, which are likely to be young stars, have been detected. We have reduced and
analyzed these original data, along with previously published X-ray observations,
in a consistent manner. Our list of {\em XMM-Newton} sources
appreciably enriches the known population of X-ray emitting stars in the
region. Then, we have collected a large catalog of photometric, spectroscopic,
and astrometric data for an even wider field, a 2.5$\times$2.5 degree area
centered on the known cluster. Most notably, the GAIA eDR3 astrometric data have
been extremely useful, albeit not per-se sufficient, to derive a reasonably
accurate census of NGC\,2264 members. Optical and IR photometry have also been
needed to clean-up the astrometrically selected candidates from field stars.
Spitzer and WISE photometry have allowed us to identify stars with a mIR excess
and therefore likely surrounded by circumstellar disks. We have derived
indication of ongoing mass accretion from narrow band H$_\alpha$ photometry, as
well as from observation of the same line with low and medium resolution
spectroscopy, these latter data being limited to the central regions.

We have assembled a catalog of more than ten-thousand candidate members,
according to at least one of several criteria, some subject to significant
field-star contamination. Out of this, we have extracted several less
contaminated samples, the most useful of which, our sample ``c'', comprises 2257 
candidate members, less than 8\% of which are expected to be contaminants,
mostly located in the outskirt of our search area. With this catalog, 
almost twice as large than previous determinations, we reach a number of
conclusions.

- We demonstrated that the optical variability of young stars, a manifestation
of magnetic activity (like the X-ray emission) and of circumstellar accretion,
is as powerful a membership criterion as X-ray emission. The effectiveness of
X-ray member selection was indeed confirmed, even in the sparse outskirts of the
region where, however, further membership confirmation is useful to reduce
field-star contamination. 

- We have defined four new substructures with respect to \citet{sun08}, based
on the surface density map of candidate members. The compact S\,Mon(C) region
may be a physically recognizable structure, being possibly older than the
surrounding region and-or having a distinctly lower fraction of stars with disks
and undergoing accretion. The embedded Cone\,(C-IR) is, instead, a recently
formed embedded subcluster. The Extended Halo and the S\,Mon(ref) region, on the
other hand, do not constitute obviously coherent structures, either
kinematically or physically.

- The identification of the population in the Extended Halo
significantly enlarges the known extent of NGC\,2264. The cluster is likely even
broader, but a less contaminated membership is needed in order to characterize
the outer population.

- NGC\,2264 is certainly not a dynamically relaxed cluster. We see signs of
ordered bulk motions such as expansion and rotation. No mass segregation,
neither spatial or in velocity-space, is clearly observed, although stars very
close to the O-type star S\,Mon might be, on average, slightly more massive than
elsewhere. Notably, we observe the collapse of filamentary structures, which may
have recently triggered new SF activity in the southern regions.

- Interpreting disk and accretion fractions as proxies of stellar ages, stars in
the southern regions are younger than those close to S\,Mon in the north. We
thus speculate that SF started $\sim$4\,Myr ago in the S\,Mon region
and extended progressively toward the south, where it is presently continuing
owing to the aforementioned collapse.

- The observed trend of the increasing disk fraction with stellar density
confirms previous suggestions that stars are preferentially born in high density
regions that later disperse as they age and dissipate their disks. The lower
disk and accretion fractions of the S\,Mon(C)$^*$ region might be due to disk
evolution, implying a slightly older age with respect to other regions, or to
the strong UV radiation of the O7 star evaporating the disks of close-by stars.
If the region were older, its high density (for the age) could be explained by a
higher-than-average density in the original cloud, or by the gravitational pull
of S\,Mon. A more precise estimation of stellar ages is required to choose
between the two possible explanations for the lower disk and accretion fractions.

Our study confirms that NGC\,2264 is one the best templates to investigate the
formation mechanisms of stars and clusters. More precise astrometric data, as
those foreseen for the future GAIA data releases, along with future
spectroscopic characterization of the new candidate members found here, will
allow us to obtain a more complete picture of the present status of the region
and of its recent evolution, to be compared with theoretical models of cloud
collapse and of star-disk evolution.

\begin{acknowledgements} 
E.F. would like to tank Francesco Damiani for suggesting to trace the position of stars back in time.
R.B. acknowledge financial support from the project PRIN-INAF 2019
``Spectroscopically Tracing the Disk Dispersal Evolution'' and the ``Preparing for Astrophysics with LSST'' Program, funded by the
Heising Simons Foundation through grant 2021-2975, and administered by Las
Cumbres Observatory.

\end{acknowledgements}

\bibliographystyle{aa} 
\bibliography{bibtex.bib} 

\begin{thebibliography}{66}
\expandafter\ifx\csname natexlab\endcsname\relax\def\natexlab#1{#1}\fi

\bibitem[{{Alencar} {et~al.}(2010){Alencar}, {Teixeira}, {Guimar{\~a}es},
  {McGinnis}, {Gameiro}, {Bouvier}, {Aigrain}, {Flaccomio}, \&
  {Favata}}]{ale10}
{Alencar}, S.~H.~P., {Teixeira}, P.~S., {Guimar{\~a}es}, M.~M., {et~al.} 2010,
  \aap, 519, A88+

\bibitem[{{Andr{\'e}} {et~al.}(2014){Andr{\'e}}, {Di Francesco},
  {Ward-Thompson}, {Inutsuka}, {Pudritz}, \& {Pineda}}]{and14}
{Andr{\'e}}, P., {Di Francesco}, J., {Ward-Thompson}, D., {et~al.} 2014, in
  Protostars and Planets VI, ed. H.~{Beuther}, R.~S. {Klessen}, C.~P.
  {Dullemond}, \& T.~{Henning}, 27

\bibitem[{{Baglin} {et~al.}(2006){Baglin}, {Auvergne}, {Boisnard}, {Lam-Trong},
  {Barge}, {Catala}, {Deleuil}, {Michel}, \& {Weiss}}]{bag06}
{Baglin}, A., {Auvergne}, M., {Boisnard}, L., {et~al.} 2006, in COSPAR Meeting,
  Vol.~36, 36th COSPAR Scientific Assembly

\bibitem[{{Ballesteros-Paredes} {et~al.}(2007){Ballesteros-Paredes}, {Klessen},
  {Mac Low}, \& {Vazquez-Semadeni}}]{bal07}
{Ballesteros-Paredes}, J., {Klessen}, R.~S., {Mac Low}, M.~M., \&
  {Vazquez-Semadeni}, E. 2007, in Protostars and Planets V, ed. B.~{Reipurth},
  D.~{Jewitt}, \& K.~{Keil}, 63

\bibitem[{{Bally} {et~al.}(1987){Bally}, {Langer}, {Stark}, \&
  {Wilson}}]{bal87}
{Bally}, J., {Langer}, W.~D., {Stark}, A.~A., \& {Wilson}, R.~W. 1987, \apjl,
  312, L45

\bibitem[{{Baraffe} {et~al.}(2015){Baraffe}, {Homeier}, {Allard}, \&
  {Chabrier}}]{bar15}
{Baraffe}, I., {Homeier}, D., {Allard}, F., \& {Chabrier}, G. 2015, \aap, 577,
  A42

\bibitem[{{Barentsen} {et~al.}(2014){Barentsen}, {Farnhill}, {Drew},
  {Gonz{\'a}lez-Solares}, {Greimel}, {Irwin}, {Miszalski}, {Ruhland}, {Groot},
  {Mampaso}, {Sale}, {Henden}, {Aungwerojwit}, {Barlow}, {Carter}, {Corradi},
  {Drake}, {Eisl{\"o}ffel}, {Fabregat}, {G{\"a}nsicke}, {Gentile Fusillo},
  {Greiss}, {Hales}, {Hodgkin}, {Huckvale}, {Irwin}, {King}, {Knigge},
  {Kupfer}, {Lagadec}, {Lennon}, {Lewis}, {Mohr-Smith}, {Morris}, {Naylor},
  {Parker}, {Phillipps}, {Pyrzas}, {Raddi}, {Roelofs}, {Rodr{\'\i}guez-Gil},
  {Sabin}, {Scaringi}, {Steeghs}, {Suso}, {Tata}, {Unruh}, {van Roestel},
  {Viironen}, {Vink}, {Walton}, {Wright}, \& {Zijlstra}}]{bar14}
{Barentsen}, G., {Farnhill}, H.~J., {Drew}, J.~E., {et~al.} 2014, \mnras, 444,
  3230

\bibitem[{{Bonito} {et~al.}(2018){Bonito}, {Hartigan}, {Venuti}, {Guarcello},
  {Prisinzano}, {Argiroffi}, {Messina}, {Johns-Krull}, {Feigelson}, {Stauffer},
  {Giannini}, {Antoniucci}, {Sciortino}, {Micela}, {Pillitteri}, {Fedele},
  {Podio}, {Damiani}, {McGehee}, {Street}, {Gizis}, {Sacco}, {Magrini},
  {Flaccomio}, {Orlando}, {Miceli}, {Stelzer}, {Fuchs}, {Chen}, {Pikuz},
  {Frasca}, {Biazzo}, {Codella}, {Pastorello}, {Alcala'}, {Covino}, {Bianchi},
  \& {Nisini}}]{bon18}
{Bonito}, R., {Hartigan}, P., {Venuti}, L., {et~al.} 2018, arXiv e-prints,
  arXiv:1812.03135

\bibitem[{{Bonito} {et~al.}(2013){Bonito}, {Prisinzano}, {Guarcello}, \&
  {Micela}}]{bon13}
{Bonito}, R., {Prisinzano}, L., {Guarcello}, M.~G., \& {Micela}, G. 2013, \aap,
  556, A108

\bibitem[{{Bonito} {et~al.}(2020){Bonito}, {Prisinzano}, {Venuti}, {Damiani},
  {Micela}, {Sacco}, {Traven}, {Biazzo}, {Sbordone}, {Masseron}, {Zwitter},
  {Gonneau}, {Bayo}, {Roccatagliata}, {Randich}, {Vink}, {Jofre}, {Flaccomio},
  {Magrini}, {Carraro}, {Morbidelli}, {Frasca}, {Monaco}, {Rigliaco}, {Worley},
  {Hourihane}, {Gilmore}, {Franciosini}, {Lewis}, \& {Koposov}}]{bon20}
{Bonito}, R., {Prisinzano}, L., {Venuti}, L., {et~al.} 2020, \aap, 642, A56

\bibitem[{{Bouvier} {et~al.}(2007){Bouvier}, {Alencar}, {Harries},
  {Johns-Krull}, \& {Romanova}}]{bou07a}
{Bouvier}, J., {Alencar}, S.~H.~P., {Harries}, T.~J., {Johns-Krull}, C.~M., \&
  {Romanova}, M.~M. 2007, Protostars and Planets V, 479

\bibitem[{{Broos} {et~al.}(2010){Broos}, {Townsley}, {Feigelson}, {Getman},
  {Bauer}, \& {Garmire}}]{bro10}
{Broos}, P.~S., {Townsley}, L.~K., {Feigelson}, E.~D., {et~al.} 2010, \apj,
  714, 1582

\bibitem[{{Buckner} {et~al.}(2020){Buckner}, {Khorrami}, {Gonz{\'a}lez},
  {Lumsden}, {Moraux}, {Oudmaijer}, {Clark}, {Joncour}, {Blanco}, {de la
  Calle}, {Hacar}, {Herrera-Fernandez}, {Motte}, {Salgado}, \&
  {Valero-Mart{\'\i}n}}]{buc20}
{Buckner}, A. S.~M., {Khorrami}, Z., {Gonz{\'a}lez}, M., {et~al.} 2020, \aap,
  636, A80

\bibitem[{{Chabrier}(2003)}]{cha03}
{Chabrier}, G. 2003, \pasp, 115, 763

\bibitem[{{Chambers} {et~al.}(2016){Chambers}, {Magnier}, {Metcalfe},
  {Flewelling}, {Huber}, {Waters}, {Denneau}, {Draper}, {Farrow}, {Finkbeiner},
  {Holmberg}, {Koppenhoefer}, {Price}, {Rest}, {Saglia}, {Schlafly}, {Smartt},
  {Sweeney}, {Wainscoat}, {Burgett}, {Chastel}, {Grav}, {Heasley}, {Hodapp},
  {Jedicke}, {Kaiser}, {Kudritzki}, {Luppino}, {Lupton}, {Monet}, {Morgan},
  {Onaka}, {Shiao}, {Stubbs}, {Tonry}, {White}, {Ba{\~n}ados}, {Bell},
  {Bender}, {Bernard}, {Boegner}, {Boffi}, {Botticella}, {Calamida},
  {Casertano}, {Chen}, {Chen}, {Cole}, {Deacon}, {Frenk}, {Fitzsimmons},
  {Gezari}, {Gibbs}, {Goessl}, {Goggia}, {Gourgue}, {Goldman}, {Grant},
  {Grebel}, {Hambly}, {Hasinger}, {Heavens}, {Heckman}, {Henderson}, {Henning},
  {Holman}, {Hopp}, {Ip}, {Isani}, {Jackson}, {Keyes}, {Koekemoer}, {Kotak},
  {Le}, {Liska}, {Long}, {Lucey}, {Liu}, {Martin}, {Masci}, {McLean}, {Mindel},
  {Misra}, {Morganson}, {Murphy}, {Obaika}, {Narayan}, {Nieto-Santisteban},
  {Norberg}, {Peacock}, {Pier}, {Postman}, {Primak}, {Rae}, {Rai}, {Riess},
  {Riffeser}, {Rix}, {R{\"o}ser}, {Russel}, {Rutz}, {Schilbach}, {Schultz},
  {Scolnic}, {Strolger}, {Szalay}, {Seitz}, {Small}, {Smith}, {Soderblom},
  {Taylor}, {Thomson}, {Taylor}, {Thakar}, {Thiel}, {Thilker}, {Unger},
  {Urata}, {Valenti}, {Wagner}, {Walder}, {Walter}, {Watters}, {Werner},
  {Wood-Vasey}, \& {Wyse}}]{cha16}
{Chambers}, K.~C., {Magnier}, E.~A., {Metcalfe}, N., {et~al.} 2016, arXiv
  e-prints, arXiv:1612.05560

\bibitem[{{Choi} {et~al.}(2016){Choi}, {Dotter}, {Conroy}, {Cantiello},
  {Paxton}, \& {Johnson}}]{cho16}
{Choi}, J., {Dotter}, A., {Conroy}, C., {et~al.} 2016, \apj, 823, 102

\bibitem[{{Cody} {et~al.}(2014){Cody}, {Stauffer}, {Baglin}, {Micela},
  {Rebull}, {Flaccomio}, {Morales-Calder{\'o}n}, {Aigrain}, {Bouvier},
  {Hillenbrand}, {Gutermuth}, {Song}, {Turner}, {Alencar}, {Zwintz},
  {Plavchan}, {Carpenter}, {Findeisen}, {Carey}, {Terebey}, {Hartmann},
  {Calvet}, {Teixeira}, {Vrba}, {Wolk}, {Covey}, {Poppenhaeger}, {G{\"u}nther},
  {Forbrich}, {Whitney}, {Affer}, {Herbst}, {Hora}, {Barrado}, {Holtzman},
  {Marchis}, {Wood}, {Medeiros Guimar{\~a}es}, {Lillo Box}, {Gillen},
  {McQuillan}, {Espaillat}, {Allen}, {D'Alessio}, \& {Favata}}]{cod14a}
{Cody}, A.~M., {Stauffer}, J., {Baglin}, A., {et~al.} 2014, \aj, 147, 82

\bibitem[{{Dahm}(2008)}]{dah08}
{Dahm}, S.~E. 2008, in Handbook of Star Forming Regions, Volume I, ed.
  B.~{Reipurth}, Vol.~4, 966

\bibitem[{{Dahm} \& {Simon}(2005)}]{dah05}
{Dahm}, S.~E. \& {Simon}, T. 2005, \aj, 129, 829

\bibitem[{{Dahm} {et~al.}(2007){Dahm}, {Simon}, {Proszkow}, \&
  {Patten}}]{dah07}
{Dahm}, S.~E., {Simon}, T., {Proszkow}, E.~M., \& {Patten}, B.~M. 2007, \aj,
  134, 999

\bibitem[{{Damiani} {et~al.}(1997){Damiani}, {Maggio}, {Micela}, \&
  {Sciortino}}]{dam97}
{Damiani}, F., {Maggio}, A., {Micela}, G., \& {Sciortino}, S. 1997, \apj, 483,
  350

\bibitem[{{Dotter}(2016)}]{dot16}
{Dotter}, A. 2016, \apjs, 222, 8

\bibitem[{{Ebeling} {et~al.}(2006){Ebeling}, {White}, \& {Rangarajan}}]{ebe06}
{Ebeling}, H., {White}, D.~A., \& {Rangarajan}, F.~V.~N. 2006, \mnras, 368, 65

\bibitem[{{Evans} {et~al.}(2018){Evans}, {Riello}, {De Angeli}, {Carrasco},
  {Montegriffo}, {Fabricius}, {Jordi}, {Palaversa}, {Diener}, {Busso},
  {Cacciari}, {van Leeuwen}, {Burgess}, {Davidson}, {Harrison}, {Hodgkin},
  {Pancino}, {Richards}, {Altavilla}, {Balaguer-N{\'u}{\~n}ez}, {Barstow},
  {Bellazzini}, {Brown}, {Castellani}, {Cocozza}, {De Luise}, {Delgado},
  {Ducourant}, {Galleti}, {Gilmore}, {Giuffrida}, {Holl}, {Kewley}, {Koposov},
  {Marinoni}, {Marrese}, {Osborne}, {Piersimoni}, {Portell}, {Pulone},
  {Ragaini}, {Sanna}, {Terrett}, {Walton}, {Wevers}, \& {Wyrzykowski}}]{eva18}
{Evans}, D.~W., {Riello}, M., {De Angeli}, F., {et~al.} 2018, \aap, 616, A4

\bibitem[{{Flaccomio} {et~al.}(2010){Flaccomio}, {Micela}, {Favata}, \&
  {Alencar}}]{fla10}
{Flaccomio}, E., {Micela}, G., {Favata}, F., \& {Alencar}, S.~P.~H. 2010, \aap,
  516, L8

\bibitem[{{Flaccomio} {et~al.}(2006){Flaccomio}, {Micela}, \&
  {Sciortino}}]{fla06}
{Flaccomio}, E., {Micela}, G., \& {Sciortino}, S. 2006, \aap, 455, 903

\bibitem[{{Flaccomio} {et~al.}(2018){Flaccomio}, {Micela}, {Sciortino}, {Cody},
  {Guarcello}, {Morales-Calder{\`o}n}, {Rebull}, \& {Stauffer}}]{fla18}
{Flaccomio}, E., {Micela}, G., {Sciortino}, S., {et~al.} 2018, \aap, 620, A55

\bibitem[{{Flewelling} {et~al.}(2020){Flewelling}, {Magnier}, {Chambers},
  {Heasley}, {Holmberg}, {Huber}, {Sweeney}, {Waters}, {Calamida}, {Casertano},
  {Chen}, {Farrow}, {Hasinger}, {Henderson}, {Long}, {Metcalfe}, {Narayan},
  {Nieto-Santisteban}, {Norberg}, {Rest}, {Saglia}, {Szalay}, {Thakar},
  {Tonry}, {Valenti}, {Werner}, {White}, {Denneau}, {Draper}, {Hodapp},
  {Jedicke}, {Kaiser}, {Kudritzki}, {Price}, {Wainscoat}, {Chastel}, {McLean},
  {Postman}, \& {Shiao}}]{fle20}
{Flewelling}, H.~A., {Magnier}, E.~A., {Chambers}, K.~C., {et~al.} 2020, \apjs,
  251, 7

\bibitem[{{Gabriel} {et~al.}(2004){Gabriel}, {Denby}, {Fyfe}, {Hoar}, {Ibarra},
  {Ojero}, {Osborne}, {Saxton}, {Lammers}, \& {Vacanti}}]{gab04}
{Gabriel}, C., {Denby}, M., {Fyfe}, D.~J., {et~al.} 2004, in Astronomical
  Society of the Pacific Conference Series, Vol. 314, Astronomical Data
  Analysis Software and Systems (ADASS) XIII, ed. F.~{Ochsenbein}, M.~G.
  {Allen}, \& D.~{Egret}, 759

\bibitem[{{Gaia Collaboration} {et~al.}(2021){Gaia Collaboration}, {Antoja},
  {McMillan}, {Kordopatis}, {Ramos}, {Helmi}, {Balbinot}, {Cantat-Gaudin},
  {Chemin}, {Figueras}, \& et~al.}]{gai21}
{Gaia Collaboration}, {Antoja}, T., {McMillan}, P.~J., {et~al.} 2021, \aap,
  649, A8

\bibitem[{{Gilmore} {et~al.}(2012){Gilmore}, {Randich}, {Asplund}, {Binney},
  {Bonifacio}, {Drew}, {Feltzing}, {Ferguson}, {Jeffries}, {Micela}, \&
  et~al.}]{gil12}
{Gilmore}, G., {Randich}, S., {Asplund}, M., {et~al.} 2012, The Messenger, 147,
  25

\bibitem[{{Guarcello} {et~al.}(2010){Guarcello}, {Damiani}, {Micela}, {Peres},
  {Prisinzano}, \& {Sciortino}}]{gua10}
{Guarcello}, M.~G., {Damiani}, F., {Micela}, G., {et~al.} 2010, \aap, 521, A18

\bibitem[{{Kenyon} \& {Hartmann}(1995)}]{ken95}
{Kenyon}, S.~J. \& {Hartmann}, L. 1995, \apjs, 101, 117

\bibitem[{{King} {et~al.}(2013){King}, {Naylor}, {Broos}, {Getman}, \&
  {Feigelson}}]{kin13}
{King}, R.~R., {Naylor}, T., {Broos}, P.~S., {Getman}, K.~V., \& {Feigelson},
  E.~D. 2013, \apjs, 209, 28

\bibitem[{{Kuhn} {et~al.}(2019){Kuhn}, {Hillenbrand}, {Sills}, {Feigelson}, \&
  {Getman}}]{kuh19}
{Kuhn}, M.~A., {Hillenbrand}, L.~A., {Sills}, A., {Feigelson}, E.~D., \&
  {Getman}, K.~V. 2019, \apj, 870, 32

\bibitem[{{Kurucz}(1993)}]{kur93a}
{Kurucz}, R. 1993, ATLAS9 Stellar Atmosphere Programs and 2 km/s grid.~Kurucz
  CD-ROM No.~13.~ Cambridge, Mass.: Smithsonian Astrophysical Observatory,
  1993., 13

\bibitem[{{Lada} \& {Lada}(2003)}]{lad03}
{Lada}, C.~J. \& {Lada}, E.~A. 2003, \araa, 41, 57

\bibitem[{{Lamm} {et~al.}(2004){Lamm}, {Bailer-Jones}, {Mundt}, {Herbst}, \&
  {Scholz}}]{lam04}
{Lamm}, M.~H., {Bailer-Jones}, C.~A.~L., {Mundt}, R., {Herbst}, W., \&
  {Scholz}, A. 2004, \aap, 417, 557

\bibitem[{{Luhman} \& {Rieke}(1999)}]{luh99b}
{Luhman}, K.~L. \& {Rieke}, G.~H. 1999, \apj, 525, 440

\bibitem[{{Mac Low} \& {Klessen}(2004)}]{mac04}
{Mac Low}, M.-M. \& {Klessen}, R.~S. 2004, Reviews of Modern Physics, 76, 125

\bibitem[{{Makidon} {et~al.}(2004){Makidon}, {Rebull}, {Strom}, {Adams}, \&
  {Patten}}]{mak04}
{Makidon}, R.~B., {Rebull}, L.~M., {Strom}, S.~E., {Adams}, M.~T., \& {Patten},
  B.~M. 2004, \aj, 127, 2228

\bibitem[{{Montillaud} {et~al.}(2019{\natexlab{a}}){Montillaud}, {Juvela},
  {Vastel}, {He}, {Liu}, {Ristorcelli}, {Eden}, {Kang}, {Kim}, {Koch}, {Lee},
  {Rawlings}, {Saajasto}, {Sanhueza}, {Soam}, {Zahorecz}, {Alina},
  {B{\"o}gner}, {Cornu}, {Doi}, {Malinen}, {Marshall}, {Micelotta}, {Pelkonen},
  {T{\'o}th}, {Traficante}, \& {Wang}}]{mon19a}
{Montillaud}, J., {Juvela}, M., {Vastel}, C., {et~al.} 2019{\natexlab{a}},
  \aap, 631, L1

\bibitem[{{Montillaud} {et~al.}(2019{\natexlab{b}}){Montillaud}, {Juvela},
  {Vastel}, {He}, {Liu}, {Ristorcelli}, {Eden}, {Kang}, {Kim}, {Koch}, {Lee},
  {Rawlings}, {Saajasto}, {Sanhueza}, {Soam}, {Zahorecz}, {Alina},
  {B{\"o}gner}, {Cornu}, {Doi}, {Malinen}, {Marshall}, {Micelotta}, {Pelkonen},
  {Viktor T{\'o}th}, {Traficante}, \& {Wang}}]{mon19b}
{Montillaud}, J., {Juvela}, M., {Vastel}, C., {et~al.} 2019{\natexlab{b}},
  \aap, 631, A3

\bibitem[{{Naranjo-Romero} {et~al.}(2022){Naranjo-Romero},
  {V{\'a}zquez-Semadeni}, \& {Loughnane}}]{nar22}
{Naranjo-Romero}, R., {V{\'a}zquez-Semadeni}, E., \& {Loughnane}, R.~M. 2022,
  \mnras, 512, 4715

\bibitem[{{Park} {et~al.}(2000){Park}, {Sung}, {Bessell}, \& {Kang}}]{par00}
{Park}, B.-G., {Sung}, H., {Bessell}, M.~S., \& {Kang}, Y.~H. 2000, \aj, 120,
  894

\bibitem[{{Peretto} {et~al.}(2006){Peretto}, {Andr{\'e}}, \&
  {Belloche}}]{per06a}
{Peretto}, N., {Andr{\'e}}, P., \& {Belloche}, A. 2006, \aap, 445, 979

\bibitem[{{Ram{\'{\i}}rez} {et~al.}(2004){Ram{\'{\i}}rez}, {Rebull},
  {Stauffer}, {Strom}, {Hillenbrand}, {Hearty}, {Kopan}, {Pravdo}, {Makidon},
  \& {Jones}}]{ram04}
{Ram{\'{\i}}rez}, S.~V., {Rebull}, L., {Stauffer}, J., {et~al.} 2004, \aj, 128,
  787

\bibitem[{{Randich} {et~al.}(2013){Randich}, {Gilmore}, \& {Gaia-ESO
  Consortium}}]{ran13}
{Randich}, S., {Gilmore}, G., \& {Gaia-ESO Consortium}. 2013, The Messenger,
  154, 47

\bibitem[{{Rebull} {et~al.}(2002){Rebull}, {Makidon}, {Strom}, {Hillenbrand},
  {Birmingham}, {Patten}, {Jones}, {Yagi}, \& {Adams}}]{reb02}
{Rebull}, L.~M., {Makidon}, R.~B., {Strom}, S.~E., {et~al.} 2002, \aj, 123,
  1528

\bibitem[{{Reipurth} {et~al.}(2004){Reipurth}, {Pettersson}, {Armond}, {Bally},
  \& {Vaz}}]{rei04a}
{Reipurth}, B., {Pettersson}, B., {Armond}, T., {Bally}, J., \& {Vaz}, L.~P.~R.
  2004, \aj, 127, 1117

\bibitem[{{Salpeter}(1955)}]{sal55}
{Salpeter}, E.~E. 1955, \apj, 121, 161

\bibitem[{{Scargle}(1982)}]{sca82}
{Scargle}, J.~D. 1982, \apj, 263, 835

\bibitem[{{Schisano} {et~al.}(2020){Schisano}, {Molinari}, {Elia},
  {Benedettini}, {Olmi}, {Pezzuto}, {Traficante}, {Brescia}, {Cavuoti}, {di
  Giorgio}, {Liu}, {Moore}, {Noriega-Crespo}, {Riccio}, {Baldeschi},
  {Becciani}, {Peretto}, {Merello}, {Vitello}, {Zavagno}, {Beltr{\'a}n},
  {Cambr{\'e}sy}, {Eden}, {Li Causi}, {Molinaro}, {Palmeirim}, {Sciacca},
  {Testi}, {Umana}, \& {Whitworth}}]{sch20}
{Schisano}, E., {Molinari}, S., {Elia}, D., {et~al.} 2020, \mnras, 492, 5420

\bibitem[{{Skrutskie} {et~al.}(2006){Skrutskie}, {Cutri}, {Stiening},
  {Weinberg}, {Schneider}, {Carpenter}, {Beichman}, {Capps}, {Chester},
  {Elias}, {Huchra}, {Liebert}, {Lonsdale}, {Monet}, {Price}, {Seitzer},
  {Jarrett}, {Kirkpatrick}, {Gizis}, {Howard}, {Evans}, {Fowler}, {Fullmer},
  {Hurt}, {Light}, {Kopan}, {Marsh}, {McCallon}, {Tam}, {Van Dyk}, \&
  {Wheelock}}]{skr06}
{Skrutskie}, M.~F., {Cutri}, R.~M., {Stiening}, R., {et~al.} 2006, \aj, 131,
  1163

\bibitem[{{Sung} {et~al.}(2008){Sung}, {Bessell}, {Chun}, {Karimov}, \&
  {Ibrahimov}}]{sun08}
{Sung}, H., {Bessell}, M.~S., {Chun}, M.-Y., {Karimov}, R., \& {Ibrahimov}, M.
  2008, \aj, 135, 441

\bibitem[{{Sung} {et~al.}(2009){Sung}, {Stauffer}, \& {Bessell}}]{sun09}
{Sung}, H., {Stauffer}, J.~R., \& {Bessell}, M.~S. 2009, \aj, 138, 1116

\bibitem[{{Tassis} \& {Mouschovias}(2004)}]{tas04}
{Tassis}, K. \& {Mouschovias}, T.~C. 2004, \apj, 616, 283

\bibitem[{{Townsley} {et~al.}(2019){Townsley}, {Broos}, {Garmire}, \&
  {Povich}}]{tow19}
{Townsley}, L.~K., {Broos}, P.~S., {Garmire}, G.~P., \& {Povich}, M.~S. 2019,
  arXiv e-prints, arXiv:1907.13126

\bibitem[{{Venuti} {et~al.}(2017){Venuti}, {Bouvier}, {Cody}, {Stauffer},
  {Micela}, {Rebull}, {Alencar}, {Sousa}, {Hillenbrand}, \&
  {Flaccomio}}]{ven17}
{Venuti}, L., {Bouvier}, J., {Cody}, A.~M., {et~al.} 2017, \aap, 599, A23

\bibitem[{{Venuti} {et~al.}(2014){Venuti}, {Bouvier}, {Flaccomio}, {Alencar},
  {Irwin}, {Stauffer}, {Cody}, {Teixeira}, {Sousa}, {Micela}, {Cuillandre}, \&
  {Peres}}]{ven14}
{Venuti}, L., {Bouvier}, J., {Flaccomio}, E., {et~al.} 2014, \aap, 570, A82

\bibitem[{{Venuti} {et~al.}(2019){Venuti}, {Damiani}, \& {Prisinzano}}]{ven19}
{Venuti}, L., {Damiani}, F., \& {Prisinzano}, L. 2019, \aap, 621, A14

\bibitem[{{Venuti} {et~al.}(2018){Venuti}, {Prisinzano}, {Sacco}, {Flaccomio},
  {Bonito}, {Damiani}, {Micela}, {Guarcello}, {Randich}, {Stauffer}, {Cody},
  {Jeffries}, {Alencar}, {Alfaro}, {Lanzafame}, {Pancino}, {Bayo}, {Carraro},
  {Costado}, {Frasca}, {Jofr{\'e}}, {Morbidelli}, {Sousa}, \& {Zaggia}}]{ven18}
{Venuti}, L., {Prisinzano}, L., {Sacco}, G.~G., {et~al.} 2018, \aap, 609, A10

\bibitem[{{Walker}(1956)}]{wal56}
{Walker}, M.~F. 1956, \apjs, 2, 365

\bibitem[{{Walter} {et~al.}(1988){Walter}, {Brown}, {Mathieu}, {Myers}, \&
  {Vrba}}]{wal88}
{Walter}, F.~M., {Brown}, A., {Mathieu}, R.~D., {Myers}, P.~C., \& {Vrba},
  F.~J. 1988, \aj, 96, 297

\bibitem[{{Weingartner} \& {Draine}(2003)}]{wei03a}
{Weingartner}, J.~C. \& {Draine}, B.~T. 2003, \apj, 589, 289

\bibitem[{{Wright} {et~al.}(2010){Wright}, {Eisenhardt}, {Mainzer}, {Ressler},
  {Cutri}, {Jarrett}, {Kirkpatrick}, {Padgett}, {McMillan}, {Skrutskie},
  {Stanford}, {Cohen}, {Walker}, {Mather}, {Leisawitz}, {Gautier}, {McLean},
  {Benford}, {Lonsdale}, {Blain}, {Mendez}, {Irace}, {Duval}, {Liu}, {Royer},
  {Heinrichsen}, {Howard}, {Shannon}, {Kendall}, {Walsh}, {Larsen}, {Cardon},
  {Schick}, {Schwalm}, {Abid}, {Fabinsky}, {Naes}, \& {Tsai}}]{wri10}
{Wright}, E.~L., {Eisenhardt}, P. R.~M., {Mainzer}, A.~K., {et~al.} 2010, \aj,
  140, 1868

\end{thebibliography}

\appendix

\section{{\em Chandra} ACIS sources}

Table \ref{tab:acis12_src} presents basic information on X-ray sources detected
on the available {\em Chandra} ACIS data for NGC\,2264 (Obs.ID: 2540, 2550,
9768, 9769, 13610, 13611, 14368, and 14369). We list: source number,
coordinates, positional uncertainties, and background-subtracted source
photons. Sources were detected with PWdetect \citep{dam97}, and analyzed using
the ACIS-EXTRACT package \citep{bro10}. The table is available in its entirety
at the CDS. A full description of the data and analysis will be provided by
Flaccomio et al. (in preparation). 

\begin{table}[!h!]
  \begin{center}
  \caption{Detected {\em Chandra} ACIS sources\label{tab:acis12_src}}
  \begin{tabular}{r r r r r }
  \hline\hline
  \expandableinput acis12_tab.tex
  \hline
  \end{tabular}
  \tablefoot{
    \tablefoottext{a}{Statitical uncertainty of position.}  
    \tablefoottext{b}{Background-subtracted source counts.}
        }
  \end{center}
\end{table}

\section{Additional diagrams used for membership}
\label{sect:modediagrams}

\begin{figure*}[!ht]
\centering
\resizebox{\textwidth}{!}{\includegraphics{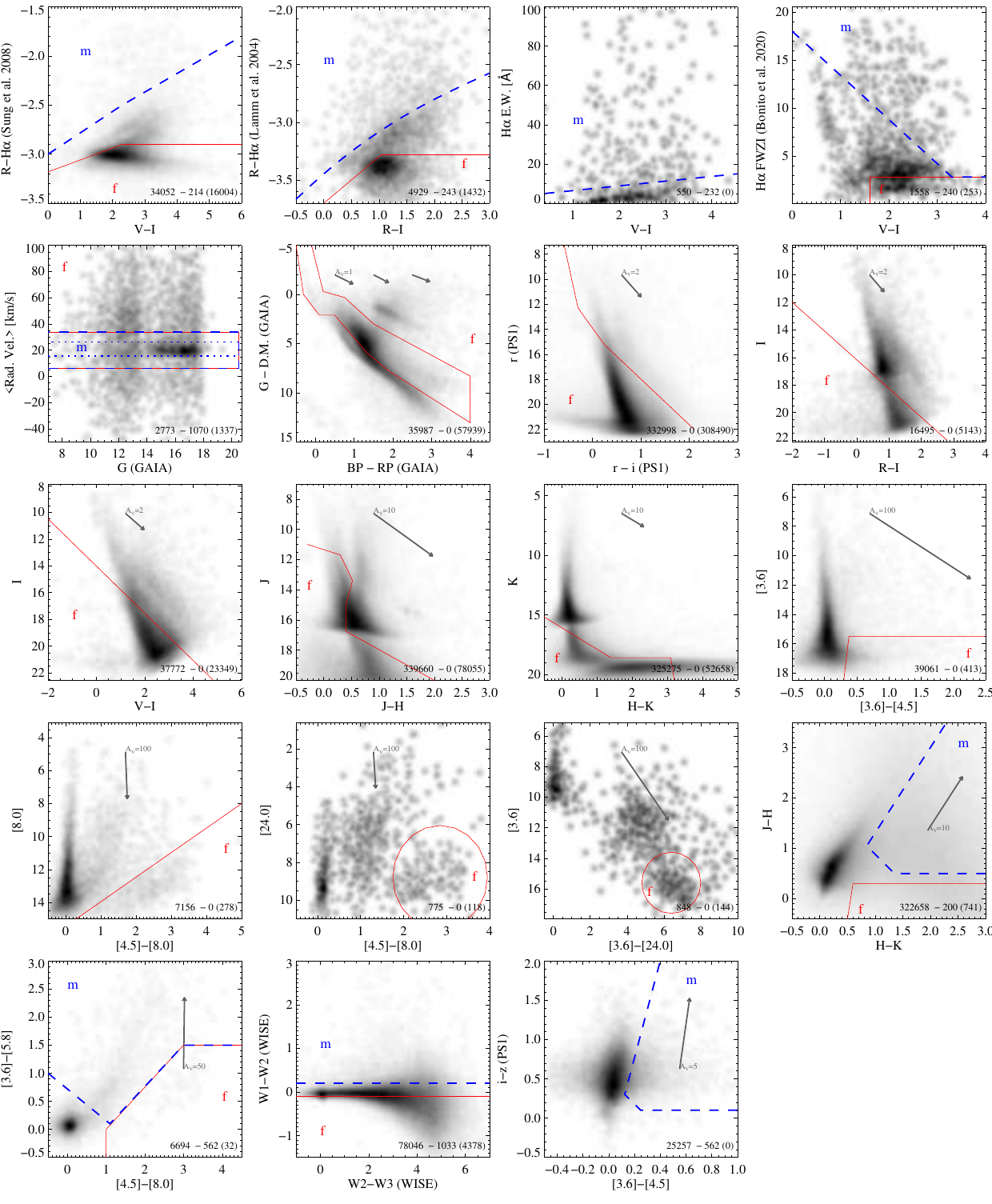}}
\caption{Nineteen additional diagrams used for membership determination in
addition to those in Fig.\,\ref{fig:GAIAdia}. All objects in our catalog are plotted. See Fig.\,\ref{fig:GAIAdia} for a description of lines and
symbols.}
\label{fig:diagrams_all}
\end{figure*}

Figure\,\ref{fig:diagrams_all} shows, for all the objects in our catalog, the 19
plots we used for selecting likely members and field objects, in addition to the
four already shown in Fig.\,\ref{fig:GAIAdia}. The following figure,
Fig.\,\ref{fig:diagrams_sampc}, shows the same plots for our final and most-used
selection of likely members (sample ``c'').

\begin{figure*}[!ht]
\centering
\resizebox{\textwidth}{!}{\includegraphics{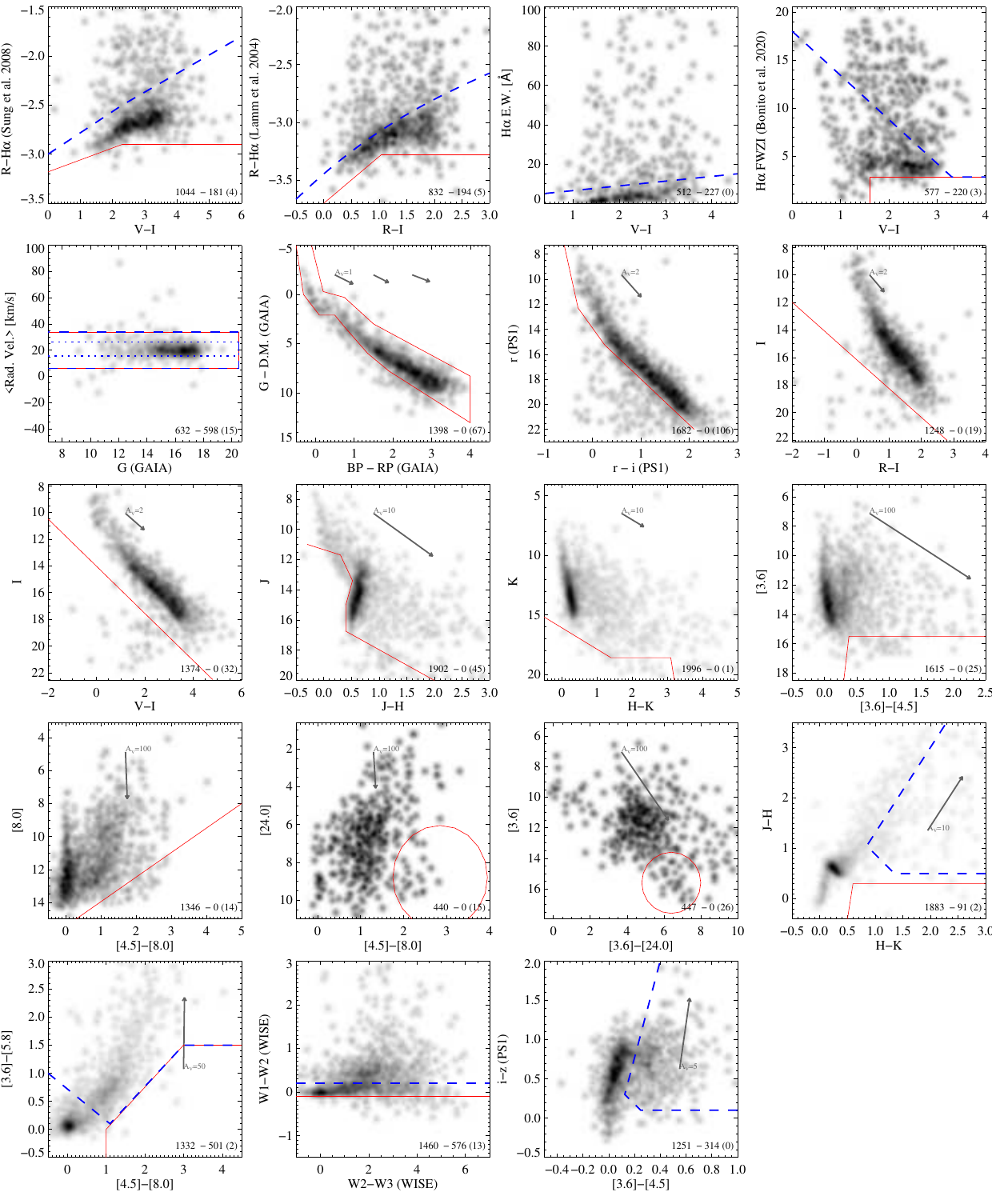}}
\caption{Same as Fig.\,\ref{fig:diagrams_all} for sample ``c''}
\label{fig:diagrams_sampc}
\end{figure*}

Within each of these figures, the first four panels in the upper row show
indicators of the strength of the H$\alpha$ line versus stellar color: the first
two use narrow-band photometric indexes similar to the IPHAS-based one in
Fig.\,\ref{fig:GAIAdia} \citep{sun08,lam04}, the third uses the spectroscopic
H$\alpha$ equivalent width (EW, from a variety of sources), and the fourth the
Full Width at Zero Intensity (FWZI) values derived by \citet{bon20} from the
spectra obtained by the GAIA-ESO survey (see their Table\,1; if more than one
spectra were available for a single source, we adopted the maximum measured
value of FWZI). For this latter plot and for those using the photometric
$H_\alpha$ indexes, we define three loci to distinguish low magnetic activity,
characteristic of old stars, strong magnetic activity, expected from young PMS
and MS stars, and strong accretion-related line emission, taken as indication of
membership. This latter locus is also defined for the spectroscopic
H$\alpha$\,EW versus V-I plot. The first panel in the second row shows
spectroscopic radial velocities versus GAIA G magnitudes. We then plot ten
different color-magnitude diagrams (CMDs). In addition to the GAIA G versus
BP-RP diagram in Fig.\,\ref{fig:GAIAdia}, we also plot the  absolute G (G -
distance modulus, DM) versus BP-RP, restricted to objects whose parallaxes have
S/N$>$5. We then show: i versus r-i\footnote{Photometry from the PS1 survey,
except for 61 stars, with missing data in our selection of the PS1 catalog.
IPHAS magnitudes are used for these stars converted to the PS1 system using the
transformations in \citet{ven19}}, I versus R-I, I versus V-I, J versus J-H, K
versus H-K, [3.6] versus [3.6]-[4.5], [8.0] versus [4.5]-[8.0], [24.0] versus
[4.5]-[8.0], and [3.6] versus [3.6]-[24.0]. In all of these CMDs we define loci
from which we select field object. Finally, we plot four color-color diagrams,
J-H versus H-K, [3.6]-[5.8] versus [4.5]-[8.0] (Spitzer), W1-W2 versus W2-W3
(WISE), and i-z versus [3.6]-[4.5] (PS1 and Spitzer). We define member loci,
based on the IR excesses produced by circumstellar disks, in all four diagrams,
and field-object loci in the first three.

\end{document}